# Data-driven fingerprint nanomechanical mass spectrometry

John E. Sader[1*], Alfredo Gomez[2], Adam P. Neumann[2], Alexander R. Nunn[2] and Michael L. Roukes[2*]

[1] *Graduate Aerospace Laboratories and Department of Applied Physics, California Institute of Technology, Pasadena, California 91125, USA*

[2] *Department of Physics, California Institute of Technology, Pasadena, California 91125, USA*

June 18, 2024

Fingerprint analysis is a ubiquitous tool for pattern recognition with applications spanning from geolocation and DNA analysis to facial recognition and forensic identification. Central to its utility is the ability to provide accurate identification without an *a priori* mathematical model for the pattern. We report a data-driven fingerprint approach for nanoelectromechanical systems mass spectrometry (NEMS-MS) that enables mass measurements of particles and molecules using complex, uncharacterized nanoelectromechanical devices of arbitrary specification. NEMS-MS is based on the frequency shifts of the NEMS vibrational modes induced by analyte adsorption. The sequence of frequency shifts constitutes a fingerprint of this adsorption, which is directly amenable to pattern matching. Two current requirements of NEMS-based mass spectrometry are: (1) *a priori* knowledge or measurement of the device mode-shapes, and (2) a mode-shape-based model that connects the induced modal frequency shifts to mass adsorption. This may not be possible for advanced NEMS with three-dimensional mode-shapes and nanometer-sized features. The advance reported here eliminates this impediment, thereby allowing device designs of arbitrary specification and size to be employed. This enables the use of advanced NEMS devices with complex vibrational modes, which offer unprecedented prospects for attaining the ultimate detection limits of nanoelectromechanical mass spectrometry.

* e-mail: jsader@caltech.edu; roukes@caltech.edu

## 1. Introduction

Mass spectrometry is used across a broad spectrum of applications, ranging from the chemical identification of compounds to the sequencing of proteins and its use in drug discovery.[1–7] These applications discriminate between analytes based on their mass-to-charge ($m/z$) ratio, often in combination with an initial chromatographic separation. In proteomic analysis, $m/z$ analysis is often complemented with information about the fragmentation patterns of proteins and protein complexes from which bioinformatics can enable identification of the original, intact analyte.[6,8–11]

In recent years, a new form of mass spectrometry has emerged—utilizing nanoelectromechanical systems (NEMS)—that circumvents these requirements by enabling direct mass measurement of the analyte, without the need for fragmentation or ionization.[12–15] Tremendous advances in this new technology have been reported over the past two decades, with the realization of routine single-molecule mass measurements and the promise of single-Dalton mass resolution.

NEMS mass spectrometry currently requires a mode-shape-based mathematical model that connects the resonant response of the sensing device to the mass of an adsorbate.[13–19] This has limited the scope of NEMS devices used to those readily amenable to mathematical modeling—e.g., using Euler-Bernoulli (E-B) elastic beam theory and thin plate/membrane theory—leading to a proliferation of studies involving cantilevered elastic beams, doubly-clamped beams, thin plate membranes and similar devices.[13–16,20,21]

For advanced NEMS devices with more complex geometries, that may exhibit three-dimensional mode shapes, numerical methods can in principle be used to compute their modes. However, small variations due to unavoidable fabrication uncertainties can alter both the shape and sequential ordering of the eigenmodes in the frequency domain (which is often used to identify the modes).[22] Miniaturization to nanoscale dimensions also obviates the use of standard optical techniques to measure these modes *in situ*. For example, recent phononic crystal resonators[23,24] (Fig. 1), and more generally advanced NEMS devices of sub-micron-sized dimensions with complex modes shapes, are difficult (perhaps not possible) to characterize using available optical techniques.[23,24] This prevents use of the existing paradigm for NEMS mass spectrometry that relies on precise knowledge of the mode shapes.

Consider a (one-dimensional) elastic beam that is currently and often used to measure the mass of an adsorbed analyte. At least two eigenmodes are needed whose fractional eigenfrequency shifts, $\Delta f_n$ where $n = 1, 2, ...$, are related to the mass, $M_{\text{analyte}}$, and position, $x$, of the adsorbed analyte, by[13–16,18]

$$\Delta f_n = -\frac{M_{\text{analyte}}}{M_{\text{device}}} \Phi_n^2(x), \qquad (1)$$

where $M_{\text{device}}$ is the device mass, it is assumed that $M_{\text{analyte}} \ll M_{\text{device}}$, $\Phi_n(x)$ is the displacement field of eigenmode $n$, and $\Delta f_n \equiv (f_n - f_n^{(0)})/f_n^{(0)}$ where $f_n$ and $f_n^{(0)}$ are the eigenfrequencies in the presence and absence of the analyte, respectively. Knowledge of the eigenmodes measured and their displacement fields, i.e., mode shapes, is central to the use of such mathematical models. These mode shapes are often approximated by theoretical idealizations, frequently leading to systematic yet unknown errors.[22] This issue presents a significant bottleneck to utilizing the full spectrum of NEMS devices that can be fabricated today and into the future.

In this article, we report a data-driven approach that circumvents these limitations by eliminating these central requirements. These are replaced with a calibration-based algorithm—termed the "learning phase" (described below)—which uses a frequency-shift "fingerprint" of several modes, to perform the mass measurement. This *fingerprint approach* opens the door to the use of nonconventional and uncharacterized NEMS devices—potentially with complex three-dimensional mode-shapes that are not amenable to experimental characterization—enabling a sole focus on NEMS device optimization for mass responsivity that is independent of device composition and geometry. This dramatically magnifies the spectrum of devices that can be employed in NEMS mass spectrometry, facilitating realization of the long-standing goal of single-Dalton mass resolution using advanced NEMS.

## 2. Fingerprint approach for mass measurement

The response of a NEMS device to an adsorbed analyte can be characterized using the sequence of fractional eigenfrequency shifts, $\Delta f_n$ ($n = 1, 2, \ldots, N$), of the eigenmodes chosen for measurement, forming the "fingerprint vector":

$$\boldsymbol{\Omega} \equiv \{\Delta f_1, \Delta f_2, \ldots, \Delta f_N\}. \tag{2}$$

For the practical case of a small and rigid analyte with $M_\text{analyte} \ll M_\text{device}$, the analyte mass affects only the magnitude of $\boldsymbol{\Omega}$ (not its direction in the $N$-dimensional *configuration space*). The overriding assumption is that the direction of the fingerprint vector is unique for mass adsorption at any single position on the device. As we shall discuss, this assumption is automatically satisfied by choosing a minimum number of eigenmodes, $N$. This eliminates the need to measure the analyte position.

The sequence of chosen eigenmodes (and hence eigenfrequencies) in Eq. (2) need not be in any particular order, and characterization of the eigenmodes is immaterial. As per traditional fingerprint analysis, a fingerprint database—denoted by the set $\mathbf{F}$ consisting of individual fingerprints, $\boldsymbol{\Omega}_\text{database}$—must be sourced for the NEMS device in question. This *learning phase* is performed experimentally through sequential adsorption of individual particles of known and identical mass, $M_\text{database}$, at random and unspecified positions on the entire device; see Fig. 1. There is no need to remove the particles between adsorption events, provided $N_\text{particle} M_\text{analyte} \ll M_\text{device}$, where $N_\text{particle}$ is the total number of adsorbed particles. The resulting locus of points from this set $\mathbf{F}$, in their $N$-dimensional configuration space, defines the required database upon which fingerprint NEMS mass spectrometry can be performed.

Importantly, the learning phase must be applied to the device of interest and cannot be used between two nominally identical devices, because of slight fabrication differences. If the device is substantially altered after performing the learning phase, then the learning phase must be repeated; adding small masses does not constitute a substantial change.

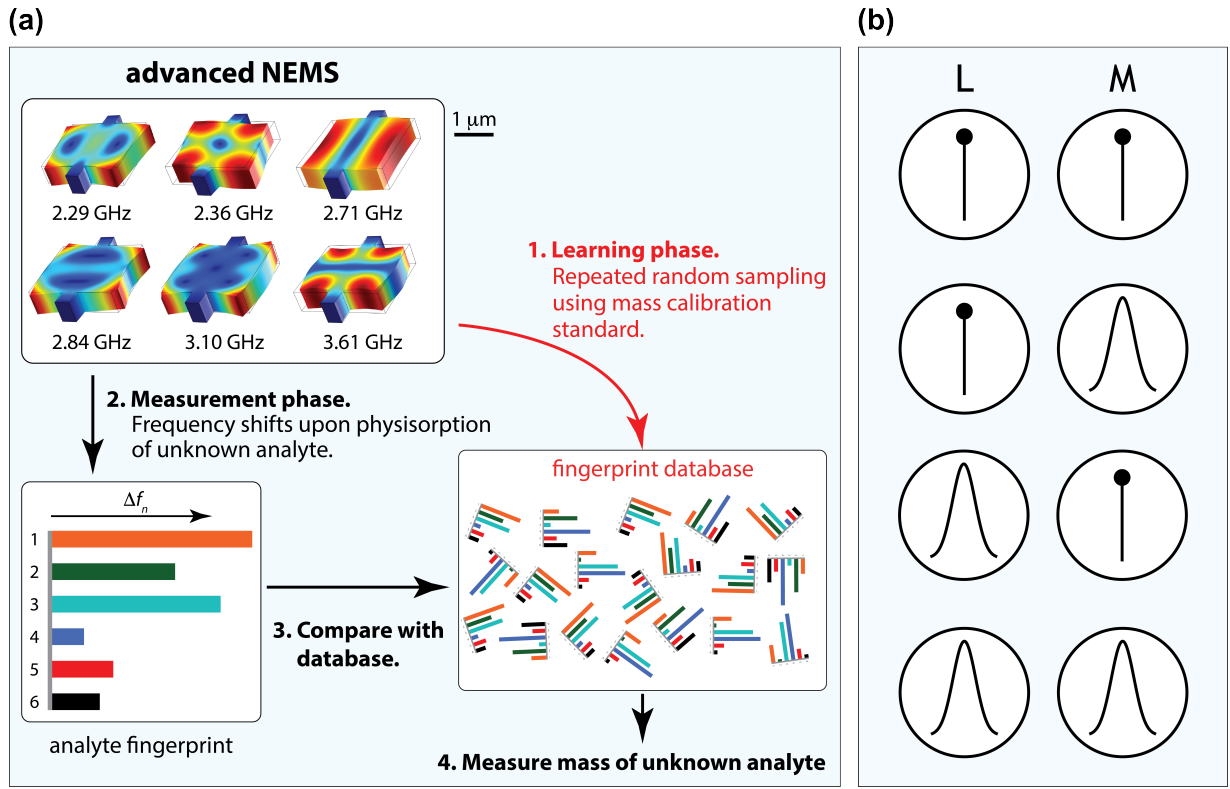

**Fig. 1 | Fingerprint approach for NEMS mass spectrometry. (a)** Schematic of fingerprint approach for NEMS mass spectrometry with an illustrative example reporting six eigenmodes of an advanced NEMS device: the defect of a phononic crystal resonator[23,24] is shown. In the *learning phase*, analytes of known mass are adsorbed at random positions on the device to determine the fingerprint database (red text: 1). In the *measurement phase*, the mass of an analyte is measured by matching its fingerprint to the fingerprint database (black text: 2, 3, 4). **(b)** Symbols to indicate the mass distributions used in the learning (L) and measurement (M) phases. Delta and Gaussian-like symbols refer to mono- and polydisperse distributions, respectively, denoting the various embodiments of the fingerprint approach. These symbols are used in subsequent figures.

In the subsequent *measurement phase*, the mass of an individual analyte, $M_{\text{analyte}}$, is determined following its random adsorption onto the surface of the NEMS device. This is achieved by measuring the analyte's fingerprint vector, $\mathbf{\Omega}_{\text{analyte}}$, and identifying the database fingerprint vector, $\mathbf{\Omega}_{\text{database}} \in \mathbf{F}$, which is most parallel to its direction. Because the direction of each fingerprint vector uniquely defines an adsorption position (see above), the unknown analyte mass can then be determined by comparing the magnitude of these fingerprint vectors, namely:

$$M_{\text{analyte}} = \frac{\|\mathbf{\Omega}_{\text{analyte}}\|}{\|\mathbf{\Omega}_{\text{parallel}}\|} M_{\text{database}}, \tag{3a}$$

where

$$\mathbf{\Omega}_{\text{parallel}} = \arg\max_{\mathbf{\Omega}_{\text{database}}} \frac{\mathbf{\Omega}_{\text{analyte}} \cdot \mathbf{\Omega}_{\text{database}}}{\|\mathbf{\Omega}_{\text{analyte}}\| \, \|\mathbf{\Omega}_{\text{database}}\|}. \tag{3b}$$

This fingerprint approach does not concern itself with the analyte's position on the NEMS device, which is an inconsequential hidden variable. It only assumes that the adsorbate weakly perturbs the dynamics of the NEMS device.

The most-parallel-vector condition arises from the discrete nature of the fingerprint database. This introduces uncertainty into the mass measurement, in addition to any other measurement noise. For analyte adsorption along a one-dimensional cantilevered beam of length, $L$, which is often used in current NEMS mass spectrometry, the resulting relative standard deviation in the measured mass is (Supplementary Information Sections A and B):

$$\text{SD } M_{\text{analyte}} = \frac{2}{N_{\text{database}}} \sqrt{\frac{2L}{x_{\min}}}, \tag{4}$$

where $N_{\text{database}}$ is the number of fingerprints in the database and $x_{\min}$ is the analyte's closest distance to the cantilever's clamped end, which is assumed to be small, i.e., the analyte position spans most of the cantilever length. For two-dimensional adsorption on an advanced NEMS device, e.g., onto the surface of a phononic crystal device,[23,24] a scaling law with respect to $N_{\text{database}}$ that varies between that of Eq. (4) and the reciprocal of $\sqrt{N_{\text{database}}}$ is expected, depending on the nature of the eigenmodes. These scaling laws establish that the "database discretization error", characterized by SD $M_{\text{analyte}}$, is systematically reduced by increasing the number of measurements, $N_{\text{database}}$, in the learning phase. Equation (4) shows that SD $M_{\text{analyte}}$ diverges if mass adsorption occurs near the clamp, and thus, such a situation must be avoided. This is not a signal-to-noise issue but a phenomenon that arises from discreteness of the fingerprint database and quasi-linear dependence of the eigenmodes near the clamp. This phenomenon can be avoided in practice by rejecting fingerprints within the measurement noise; see Supplementary Information Section C.

## 3. Phononic crystal device

To demonstrate the fingerprint approach, we first report synthetic numerical simulations using a Monte Carlo algorithm that implements the above-described procedure on an advanced NEMS device: a phononic crystal resonator.[23,24] Analogous results for a cantilevered beam are given in Supplementary Information Section D. The resonating adsorption platform (the "defect site"[23,24]) of these gigahertz frequency phononic crystal devices are approximately one micron in size which prohibits conventional optical characterization of their eigenmodes.[22] Nonidealities in fabrication can distort these eigenmodes and their ordering in frequency, given these resonant frequencies are closely spaced; see Fig. 1. Moreover, the eigenmodes undergo shear (in-plane) deformation which further complicates mode-shape measurement; see Fig. 1.

The Monte Carlo algorithm places particles, in both the learning and measurement phases, randomly onto the device surface while calculating the resulting fingerprints from the mode-shapes. This uses the two-dimensional generalization of Eq. (1)—because the analyte adsorbs to a surface of the device—and simulates an actual experiment where adsorption positions are not recorded. As per Fig. 1, the fingerprint database is generated in the learning phase by adsorbing (calibrated) particles of a single specified mass, $M_{\text{database}}$, numerous times at random positions on the device surface; 1,000 particles are used in this demonstration. The resulting fingerprint databases, generated using the first 3 and 4 eigenmodes within the band gap of the device,[23,24] are shown in Figs. 2(a, d). This learning phase primes the fingerprint approach, which can then be used in the measurement phase to determine the unknown mass of an analyte particle.

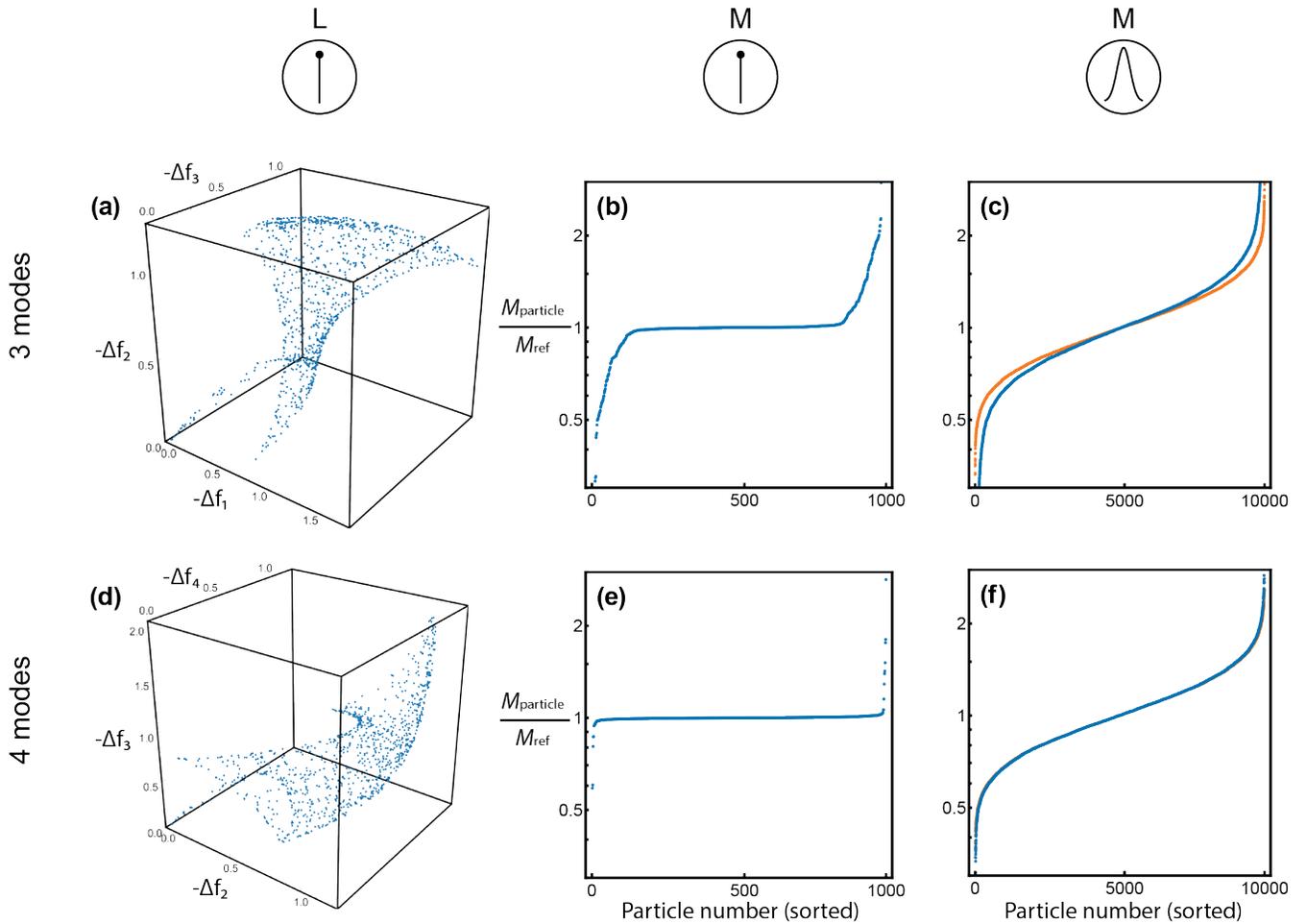

**Fig. 2 | Numerical simulations of a phononic crystal device.**[23,24] Fingerprint approach using the first three (2.29, 2.36 and 2.71 GHz) and four (2.29, 2.36, 2.71 and 2.84 GHz) eigenmodes of a phononic crystal resonator (in its measurement band gap) illustrated in Fig. 1. **(a, d)** Learning phase showing fingerprint databases with $N_{\text{database}} = 1,000$ fingerprints, which are used in all measurement phases. **(b, e)** Measurement phases for 1,000 identical analyte particles, shown as an ordered density plot. **(c, f)** Measurement phases for 10,000 different analyte particles, whose masses obey a log-normal distribution. Orange curve is the true mass distribution. Blue dots (which appear as a single curve) are the measured masses using the fingerprint approach.

Figure 2(b, e) gives the measurement phase of the fingerprint approach using (additional) 1,000 analyte particles. The analyte particles are of identical mass to those used in the learning phase. Changing the analyte mass gives identical plots in the measurement phase, except that they are rescaled by the relative mass change due to linearity of Eq. (3a). The particle mass, averaged over all measurements, is accurately determined in the measurement phase. This is regardless of the number

of eigenmodes used, though using 4 eigenmodes gives substantially less variance. The displayed sorted density plots highlight this feature: using 3 eigenmodes gives the correct mass with a standard deviation of ~40% over all measurements, while use of 4 eigenmodes exhibits a standard deviation of only ~4%.

The poorer accuracy in the 3-eigenmode measurement phase data is due to the multivalued nature of its fingerprint database. While some fingerprints are uniquely defined by their direction in the configuration space, others are not, i.e., there exist multiple branches of solution. This violates the overriding assumption of uniqueness in the fingerprint vector (see above), which is guaranteed when 4 or more eigenmodes are used for two-dimensional adsorption on the physical surfaces of NEMS devices; see Supplementary Information Section D and for adsorption on a one-dimensional device.

Figures 2(c, f) give mass measurements (using the same fingerprint databases shown in Figs. 2(a, d)) for 10,000 analyte particles obeying a log-normal mass distribution in the measurement phase only. More analyte particles are used here in the measurement phase to reduce uncertainty in the sorted density plots and ensure an accurate comparison. The above-mentioned multivalued nature of the 3-eigenmode database in the learning phase is apparent in Fig. 2(c), with a visible difference in the measured and true mass distributions. Increasing the number of eigenmodes to four immediately eliminates any visible difference, see Fig. 2(f), which is restricted to the above-mentioned database discretization error.

**4. Experimental validation**

The fingerprint approach is validated using experimental measurements on two current NEMS mass spectrometry devices: (1) a doubly-clamped elastic beam operating in vacuum, and (2) a suspended microchannel resonator (SMR) for buoyant mass measurements in liquid. The purpose is to demonstrate and benchmark the fingerprint approach on actual experimental data. Nominally identical particles are used in both the learning and measurement phases, so that the measured masses are implicitly calibrated; use of different particles would require independent calibration of their masses and generate additional uncertainty in this benchmark study.

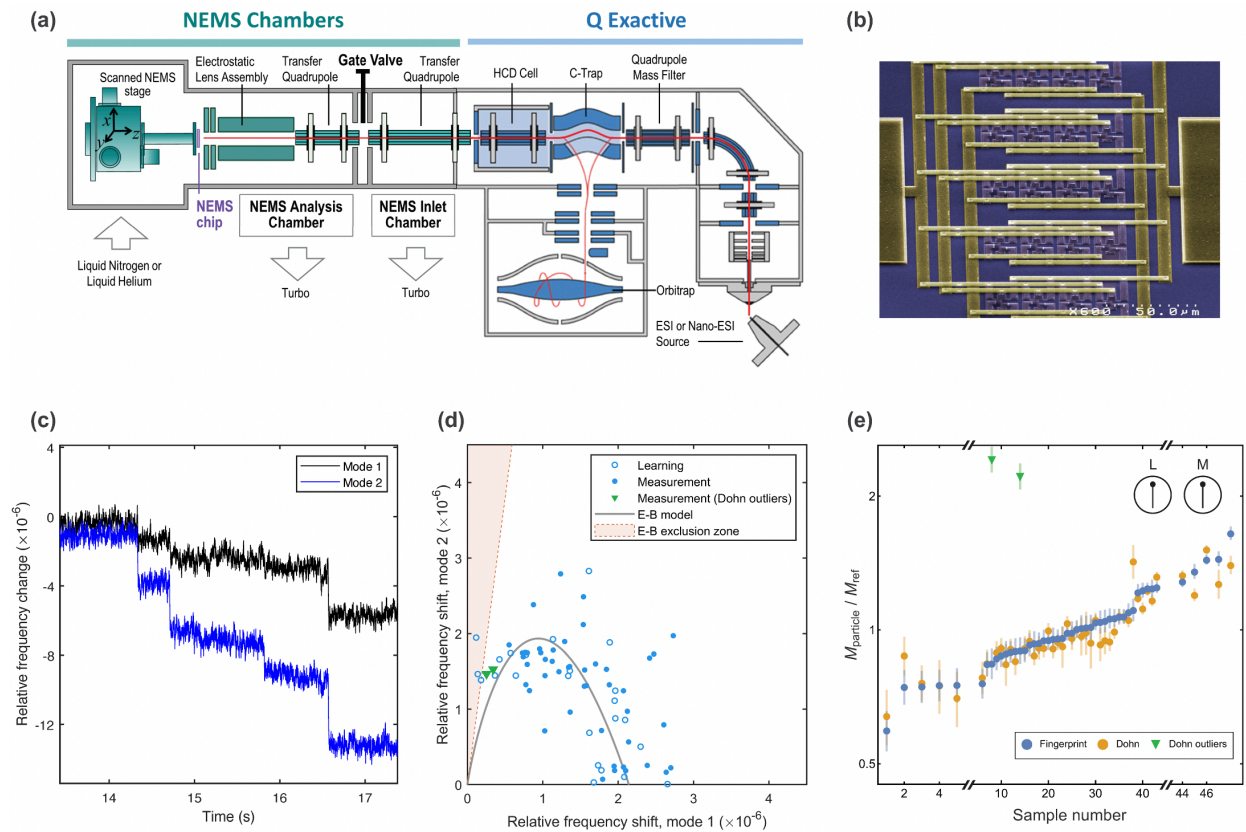

**Fig. 3 | Mass measurements of GroEL molecules with a doubly-clamped NEMS beam in high vacuum. (a)** Architecture of the Hybrid Q Exactive-NEMS System that delivers intact proteins to the Orbitrap chamber for analysis of mass-to-charge ratio and then onto the NEMS for single molecule analysis. **(b)** SEM image of a 20-device array of doubly-clamped beams showing their metallization layers, AlSi (colorized in yellow), used to interconnect the electrical connections of each resonator. **(c)** As GroEL molecules physisorb to a single NEMS resonator, the resonant frequency of each tracked flexural eigenmode abruptly shifts. Sub-figures (a) – (c) taken from Ref. 26. **(d)** All 72 measured fingerprints of the first two eigenmodes in their configuration space: [open circles] 24 fingerprints in the learning phase; [closed circles] 48 fingerprints in the measurement phase; [triangles] 2 fingerprints that are outliers of the Dohn method; [solid line] best fit to the E-B model; [shaded region] exclusion zone of E-B model where the Dohn method does not apply. **(e)** Measured mass of 48 particles using: [blue dots] the measurement phase of the fingerprint approach; [orange dots] the Dohn method; [triangles] the Dohn method which does not give a converged solution.

*A. Doubly-clamped beam measurements in vacuum*

We first apply the fingerprint approach to measure the mass of individual proteins physisorbed onto the surface of a doubly-clamped NEMS beam in high vacuum. These macromolecules are of identical (fixed) mass to within a degree of high precision; modulated only by the binding of hydrogen, water molecules etc. A hybrid Orbitrap-NEMS system illustrated in Fig. 3(a) is used to perform single-molecule nanomechanical mass measurements of E. coli GroEL chaperonin, a noncovalent complex consisting of 14 identical subunits; its theoretical molecular mass is 800.7664 kDa for which independent measurements have confirmed a value of 800.7822±0.0236 (SD) kDa.[25–27] GroEL is pre-selected using the quadrupoles of the orbitrap system, ensuring only intact GroEL molecules are delivered to the NEMS. A 20-device NEMS array of doubly-clamped beams (Fig. 3(b)) is used to localize the focal point of the ion beam. Subsequently, the smallest NEMS device in the array with the best mass resolution (length of 7 μm) is operated in isolation. Details concerning the operation and fabrication of this array are provided in Refs. 25–28.

Individual molecular adsorption events of intact GroEL molecules abruptly shift the resonant frequencies of the first two flexural eigenmodes of the smallest NEMS device; see experimental data in Fig. 3(c). The tracked frequency data collected for each flexural eigenmode are time series, with jump events due to molecular adsorption. These jump events are detected and analyzed using a statistical algorithm described in Ref. 29. This produces a dataset consisting of pairs of frequency shifts (for flexural eigenmodes 1 and 2 of the device) for each molecular adsorption event, each of which is a fingerprint.

The fingerprint approach is used to analyze this measured fingerprint dataset to recover the mass of each molecule. The result is compared to a conventional mode-shape-based method reported by Dohn *et al.*[18] (termed, the "Dohn method"), which fits the relative frequency shifts of multiple eigenmodes—measured at a single particle position—to E-B theory using a weighted least-squares approach. Uncertainty in the mass measurements is reported, allowing for a robust comparison. In the learning phase, the fingerprint database is generated from the first 24 entries of the overall frequency shift dataset. The remaining 48 entries in the frequency shift dataset are then analyzed in the measurement phase of the fingerprint approach, to determine the masses of their corresponding GroEL molecules. The Dohn method is applied directly to each entry of the fingerprint database, with the median of the resulting 24 measured masses used as the mass reference in the measurement phase.

Figure 3(d) shows a plot of all fingerprints, for the learning and measurement phases, in their configuration space. Figure 3(d) also shows the exclusion zone of E-B theory which is not accessible by the Dohn method, i.e., no particle mass can be recovered from these positions in the configuration space. In contrast, the fingerprint approach has no such limitation and naturally handles this situation (which may naturally arise due to frequency noise). Comparison of the measured masses of each of the above-described 48 GroEL molecules using the fingerprint approach and the Dohn method is reported in Fig. 3(e). Excellent agreement is observed using these independent approaches. This constitutes an experimental validation of the proposed fingerprint approach for a widely-used NEMS mass spectrometry set up: physisorption of a small analyte onto the surface of a NEMS device, showing that the fingerprint approach can be used with confidence.

Next, we apply the fingerprint approach to another experimental configuration, that has established capability in performing highly sensitive measurements in liquid.

*B. Suspended microchannel resonator measurements in liquid*

The suspended microchannel resonator (SMR) consists of a cantilevered beam with a microfluidic channel embedded in its interior, through which an analyte immersed in liquid can flow;[21,30–32] see Fig. 4(a). As the analyte passes through the microfluidic channel of the device, the resonant frequencies of its multiple flexural eigenmodes are simultaneously recorded in real-time; see Supplementary Information Section E. Because the analyte is immersed in liquid, the resonant-frequency time series depend on the buoyant mass of the analyte.[21,30–32] Experimental data is taken from Ref. 33. This data consists of frequency-shift time series as a single NIST-tracible polystyrene particle (ThermoFisher 4016A) passes through the SMR, for flexural eigenmodes 2 to 6. A total of 341 individual polystyrene particles, and hence time series, are used.

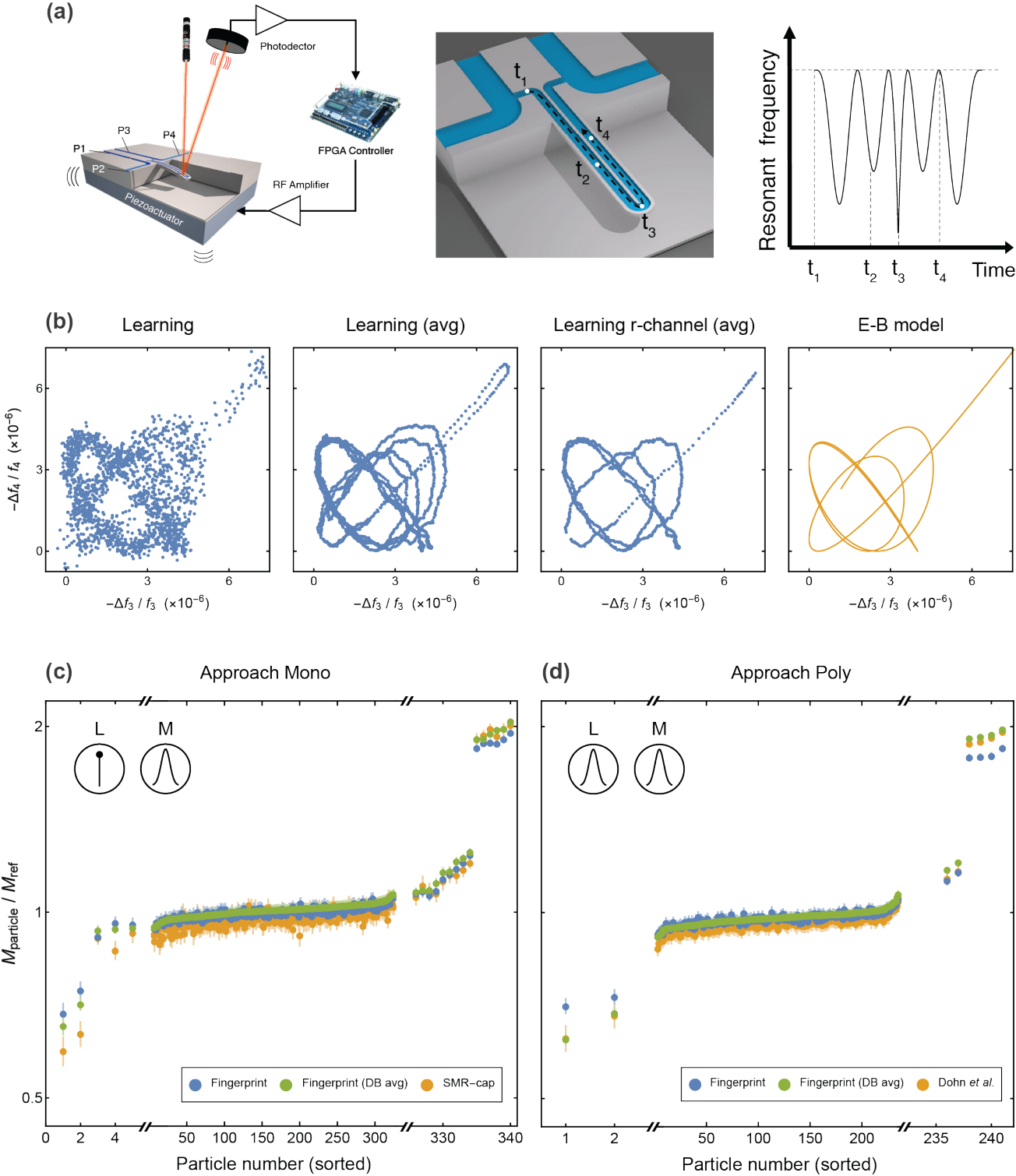

**Fig. 4 | Application of fingerprint approach to SMR mass measurements. (a)** Schematic of a suspended microchannel resonator (SMR) with actuation/detection scheme (left); interior of SMR as a particle traverses its microfluidic channel (middle); and corresponding resonant frequency time series (right). Taken from Ref. 33. **(b)** Fingerprint databases for the learning phase generated from: [Learning] the measured frequency-shift time series of a single particle; [Learning (avg)] the same time series processed using a 20-point moving average; [Learning r-channel (avg)] the same averaged time series only in the right-hand microfluidic channel; [E-B model] modeling using E-B theory. **(c, d)** Measured relative particle mass in the measurement phases of Approaches Mono and Poly, using: [Fingerprint] the unaveraged fingerprint database; [Fingerprint (DB avg)] the averaged fingerprint database; [SMR-*cap*] the conventional measurement approach with no averaging of the time series; [Dohn *et al.*] the Dohn method.[18] The reference mass, $M_{\text{ref}}$, for SMR-*cap* is the median of all mass measurements of the particle used in the learning phase; for the Dohn method it is the median of all mass measurements generated using each element of the fingerprint database. The reference mass, $M_{\text{ref}}$, for the fingerprint approach is implicitly specified by its database.

SMR measurements are conventionally analyzed by fitting the measured frequency-shift time series dataset of each eigenmode to the squared magnitudes of the mode-shapes from E-B theory;[21,30–32] such a times series is schematically illustrated in the right-hand panel of Fig. 4(a). This SMR conventional analysis procedure (termed, SMR-*cap*) extracts the analyte-to-SMR mass ratio from each eigenmode, which can then be averaged to obtain the required particle mass.

SMR-*cap* and the Dohn method use models based on E-B theory to extract the buoyant mass of a particle from the measured data, and they both require precise knowledge of the eigenmodes measured and their mode shapes. The fingerprint approach replaces these mode-shaped-based models and required eigenmode knowledge with a subset of the measured data taken from specific particle(s) in its learning phase. The buoyant masses of the remaining particles are then determined in the measurement phase using a two-step procedure. This first involves application of Eq. (3b), then Eq. (3a), to the measured fingerprint at each discrete time as the particle traverses the microfluidic channel. Thousands of buoyant mass measurements for each individual particle are thus determined, from which their median buoyant mass and 95% C.I. are reported.

We benchmark the results of the fingerprint approach to the mode-shape-based approaches of SMR-*cap* and the Dohn method. Two separate embodiments of the fingerprint algorithm are employed. The first embodiment (termed, Approach Mono) uses a *single* particle in the learning phase whereas the second embodiment (termed, Approach Poly) uses *multiple* particles (with a distribution of masses). The measurement phase of Approach Mono is compared to SMR-*cap* whereas that of Approach Poly is compared to the Dohn method. Approach Mono represents the idealized implementation of the fingerprint method, whereby the same single particle is placed at different positions along the device. Approach Poly is what would occur more often in a general NEMS measurement, whereby nominally identical but different particles are placed at different positions along the device.

**Approach Mono:** Here, we choose the measured time series dataset of one particle only for the learning phase, to generate the fingerprint database. This is possible in an SMR measurement because a single particle samples many positions along the entire device, as the particle passes continuously through its microfluidic channel. Each mass of the remaining 340 particles is determined in the measurement phase.

*Learning phase:* Figure 4(b) shows the resulting fingerprint database (using one particle only), consisting of thousands of individual fingerprints that are generated in the learning phase. Results are shown for only two out of the five flexural eigenmodes measured, for illustrative purposes. In addition, Fig. 4(b) shows a fingerprint database generated by smoothing the same time series using a simple 20-point moving average. This averaging process visually sharpens the database in its configuration space, the features of which are otherwise obscured by noise. Figure 4(b) also gives the theoretical E-B theory prediction for this subset of the fingerprint database, which displays excellent agreement with the (averaged) measurements of the same database.

*Measurement phase:* Figure 4(c) gives the mass measured with SMR-*cap*, and in the measurement phase of the fingerprint approach using the unaveraged and averaged databases. While most particle buoyant masses closely align with that of the reference particle used to generate the learning phase—giving a relative mass of one—there are several strong outliers. Coincidence of buoyant mass measurements from the fingerprint approach with those of SMR-*cap* shows that these outliers are not erroneous measurements but represent real mass differences, which the fingerprint approach naturally handles.

**Approach Poly:** The learning phase of Approach Poly is generated by randomly sampling a single fingerprint from SMR time series measurements on 100 different particles; this gives a fingerprint database with 100 fingerprints. These particles have a narrow mass distribution with ThermoFisher specifying a standard deviation of 3%. In the measurement phase, the fingerprint approach is applied to all fingerprints from each time series of the remaining 241 particles. The median of the measured masses of each of these particles and its 95% C.I. are reported in Fig. 4(d). This mass dataset is compared to that obtained independently using the (multimode) Dohn method, which is an equivalent mode-shape-based approach. Equation (4) shows that the resulting database discretization error in the measurement phase of the fingerprint approach is small and in the single-percentile range. Good agreement is observed between these independent, yet complementary, approaches. We also find that using the averaged fingerprint database enhances the level of agreement.

## 5. Features of the fingerprint approach

The principal advantage of the fingerprint approach over existing mode-shape-based methods is that it does not require or attempt to estimate the device mode shapes, i.e., no knowledge of the mode shapes is needed. Instead, its learning phase uses a (known) mass standard to characterize the device response to (unknown) mass adsorption. This is particularly advantageous when the mode shapes of the device in question cannot be easily characterized or identified from its measured frequency spectrum. This is the case for the phononic crystal devices studied above, whose frequency spectrum is dense and does not permit definitive identification of the corresponding mode shapes in the presence of non-idealities. As discussed, inevitable fabrication errors can lead to small variations in the device shape or material properties, which can alter the expected frequency ordering of its vibrational modes.

Moreover, all mode-shape-based methods require the device to be characterized for their implementation, e.g., by independently measuring the device mass, stiffness, geometry and potentially the mode shapes (if not available via theoretical/numerical calculation). This is replaced with the learning phase in the fingerprint approach. Consequently, the fingerprint approach can directly measure mass by comparison to a primary standard, e.g., GroEL biomolecules (used in Section 4A), in contrast to mode-shape-based methods where device characterization/modelling is a necessary error-inducing intermediatory.

The learning phase of the fingerprint method introduces a new uncertainty into NEMS-MS in the form of its database discretization error, discussed in Section 2 and Supplementary Information Sections A and B. Importantly, this error is small (see above) and can be systematically reduced by increasing the number of calibration measurements in the learning phase. This contrasts with conventional mode-shape-based methods where the uncertainty or error in the mode shapes is fixed, e.g., due to unknown tension in the device and/or variations in its clamping and material properties. Another source of error in the fingerprint approach arises from uncertainty or variation in the mass calibration standard used in its learning phase. This can be mitigated by using biomolecules whose mass is dictated by their molecular structure, e.g., the molecular mass of GroEL is precisely 800.7664 kDa.[25] The use of GroEL as a mass standard was demonstrated in this study, along with NIST-tracible polystyrene particles which were used to examine the effect of polydispersity in the chosen mass standard.

As is the case for current NEMS-MS approaches, the presence of degenerate modes in the device may prove to be problematic.[34] However, the required multiple particle adsorption events in the learning phase can eliminate the effect of any modal degeneracy in the present fingerprint approach.

The effect of mode degeneracy in NEMS-MS thus represents an interesting area of future investigation for both the fingerprint approach and conventional NEMS-MS analysis.

## 6. Conclusions and outlook

NEMS mass spectrometry presents novel advantages over standard, widely used techniques, particularly in its ability to measure mass without the need for molecular ionization. A significant impediment to its wider application is the requirement for accurate knowledge of the NEMS device mode shapes, which are rarely if ever measured (or verified) in practice. Advanced NEMS devices can have three-dimensional mode-shapes of nanometer size, which further complicates matters. The fingerprint approach reported here circumvents this mode-shape requirement, thereby permitting the use of NEMS devices of arbitrary geometry and specification. The current requirement for calibration of the device mass is also eliminated. This now enables the use of advanced NEMS devices, especially those with arbitrarily complex mode-shapes, which can be designed to achieve exceptionally low noise and high responsivity. Among contemporary examples are phononic bandgap NEMS devices[23,24] (studied theoretically here in Section 3) and those engineered to have ultralow energy dissipation[35] which offer the prospect of attaining single-Dalton mass resolution. The three-dimensional eigenmodes of these devices are generally vastly different and densely packed in the frequency domain as compared with the widely used one-dimensional beam modes; as demonstrated in Fig. 1. This spectral complexity can preclude definitive identification of the eigenmodes solely from the measured frequency spectra of such devices, as conventionally performed. The proposed fingerprint approach naturally accommodates such dimensionality and complexity, permitting the use of arbitrarily complex NEMS resonators. This advance dramatically expands the capabilities of NEMS mass spectrometry.

**Data availability**

The data supporting the findings of this study are available from the corresponding author upon reasonable request.

**Code availability**

The code used to generate the findings of this study are available via Zenodo using the identifiers; 10.5281/zenodo.10211934, 10.5281/zenodo.10211936.

**Acknowledgments**

The authors thank J. F. Collis and S. R. Manalis for useful discussions. The authors gratefully acknowledge support from the Wellcome Leap Foundation through its Delta Tissue program.

**Author contributions**

J.E.S. proposed the fingerprint method for NEMS mass spectrometry and A.G. introduced the most parallel-vector approach. A.P.N. developed the noise thresholding methodology and A.R.N. contributed to the variance calculation of the positional discrepancy. J.E.S. performed the mass uncertainty calculation, analyzed the SMR data and wrote the paper. M.L.R. supervised the project and provided its overall direction. All authors contributed to the data analysis and ensuing discussions regarding its interpretation, as well as editing of the manuscript.

**References**

1. Glish, G. L. & Vachet, R. W. The basics of mass spectrometry in the twenty-first century. *Nat. Rev. Drug Discov.* **2,** 140–150 (2003).
2. Rivier, L. Criteria for the identification of compounds by liquid chromatography–mass spectrometry and liquid chromatography–multiple mass spectrometry in forensic toxicology and doping analysis. *Anal. Chim. Acta* **492,** 69–82 (2003).
3. Mann, M., Hendrickson, R. C., Pandey, A., & others. Analysis of proteins and proteomes by mass spectrometry. *Annu. Rev. Biochem.* **70,** 437–473 (2001).
4. Domon, B. & Aebersold, R. Mass spectrometry and protein analysis. S*cience* **312,** 212–217 (2006).
5. Korfmacher, W. A. Foundation review: Principles and applications of LC-MS in new drug discovery. *Drug Discov. Today* **10,** 1357–1367 (2005).

6. Yates, J. R., Ruse, C. I. & Nakorchevsky, A. Proteomics by mass spectrometry: approaches, advances, and applications. *Annu. Rev. Biomed. Eng.* **11,** 49–79 (2009).

7. Biemann, K. Mass Spectrometry. *Annu. Rev. Biochem.* **32,** 755–780 (1963).

8. Bantscheff, M., Schirle, M., Sweetman, G., Rick, J. & Kuster, B. Quantitative mass spectrometry in proteomics: a critical review. *Anal. Bioanal. Chem.* **389,** 1017–1031 (2007).

9. Bantscheff, M., Lemeer, S., Savitski, M. M. & Kuster, B. Quantitative mass spectrometry in proteomics: critical review update from 2007 to the present. *Anal. Bioanal. Chem.* **404,** 939–965 (2012).

10. Käll, L., Canterbury, J. D., Weston, J., Noble, W. S. & MacCoss, M. J. Semi-supervised learning for peptide identification from shotgun proteomics datasets. *Nature Methods* **4,** 923–925 (2007).

11. Locard-Paulet, M., Bouyssié, D., Froment, C., Burlet-Schiltz, O. & Jensen, L. J. Comparing 22 popular phosphoproteomics pipelines for peptide identification and site localization. *J. Proteome Res.* **19,** 1338–1345 (2020).

12. Naik, A. K., Hanay, M. S., Hiebert, W. K., Feng, X. L. & Roukes, M. L. Towards single-molecule nanomechanical mass spectrometry. *Nature Nanotech.* **4,** 445–450 (2009).

13. Hanay, M. S., Kelber, S., Naik, A. K., Chi, D., Hentz, S., Bullard, E. C., Colinet, E., Duraffourg, L. & Roukes, M. L. Single-protein nanomechanical mass spectrometry in real time. *Nature Nanotech.* **7,** 602–608 (2012).

14. Sader, J. E., Hanay, M. S., Neumann, A. P. & Roukes, M. L. Mass Spectrometry Using Nanomechanical Systems: Beyond the Point-Mass Approximation. *Nano Lett.* **18,** 1608–1614 (2018).

15. Sage, E., Brenac, A., Alava, T., Morel, R., Dupré, C., Hanay, M. S., Roukes, M. L., Duraffourg, L., Masselon, C. & Hentz, S. Neutral particle mass spectrometry with nanomechanical systems. *Nat. Commun.* **6,** 1–5 (2015).

16. Hanay, M. S., Kelber, S. I., O'Connell, C. D., Mulvaney, P., Sader, J. E. & Roukes, M. L. Inertial imaging with nanomechanical systems. *Nature Nanotech.* **10,** 339–344 (2015).

17. Ruz, J., Malvar, O., Gil-Santos, E., Calleja, M. & Tamayo, J. Effect of particle adsorption on the eigenfrequencies of nano-mechanical resonators. *J. Appl. Phys.* **128,** 104503 (2020).

18. Dohn, S., Svendsen, W., Boisen, A. & Hansen, O. Mass and position determination of attached particles on cantilever based mass sensors. *Rev. Sci. Instrum.* **78,** 103303 (2007).

19. Dominguez-Medina, S., Fostner, S., Defoort, M., Sansa, M., Stark, A.-K., Halim, M. A., Vernhes, E., Gely, M., Jourdan, G., Alava, T., Boulanger, P., Masselon, C. & Hentz, S. Neutral mass spectrometry of virus capsids above 100 megadaltons with nanomechanical resonators. *Science* **362,** 918–922 (2018).

20. Atalaya, J., Kinaret, J. M. & Isacsson, A. Nanomechanical mass measurement using nonlinear response of a graphene membrane. *EPL Europhys. Lett.* **91,** 48001 (2010).

21. Godin, M., Bryan, A. K., Burg, T. P., Babcock, K. & Manalis, S. R. Measuring the mass, density, and size of particles and cells using a suspended microchannel resonator. *Appl. Phys. Lett.* **91,** 123121 (2007).

22. Sartori, A. F., Belardinelli, P., Dolleman, R. J., Steeneken, P. G., Ghatkesar, M. K. & Buijnsters, J. G. Inkjet-Printed High-Q Nanocrystalline Diamond Resonators. *Small* **15,** 1803774 (2019).

23. Arrangoiz-Arriola, P. & Safavi-Naeini, A. H. Engineering interactions between superconducting qubits and phononic nanostructures. *Phys. Rev. A* **94,** 063864 (2016).

24. Wollack, E. A., Cleland, A. Y., Arrangoiz-Arriola, P., McKenna, T. P., Gruenke, R. G., Patel, R. N., Jiang, W., Sarabalis, C. J. & Safavi-Naeini, A. H. Loss channels affecting lithium niobate phononic crystal resonators at cryogenic temperature. *Appl. Phys. Lett.* **118,** 123501 (2021).

25. Rose, R. J., Damoc, E., Denisov, E., Makarov, A. & Heck, A. J. High-sensitivity Orbitrap mass analysis of intact macromolecular assemblies. *Nature Methods* **9,** 1084–1086 (2012).

26. Neumann, A. P. Towards Single Molecule Imaging Using Nanoelectromechanical Systems. Dissertation (Ph.D.), California Institute of Technology, DOI:10.7907/n4ap-7h91 (2020).

27. Sage, E., Sansa, M., Fostner, S., Defoort, M., Gély, M., Naik, A. K., Morel, R., Duraffourg, L., Roukes, M. L. & Alava, T. Single-particle mass spectrometry with arrays of frequency-addressed nanomechanical resonators. *Nature Comm.* **9,** 3283 (2018).

28. Neumann, A. P., Sage, E., Boll, D., Reinhardt-Syzba, M., Fon, W., Masselon, C., Hentz, S., Sader, J. E., Makarov, A. & Roukes, M. L. A hybrid Orbitrap-NEMS instrument for real-time single-molecule analysis of intact proteins, submitted for publication.

29. Neumann, A. P., Gomez, A., Nunn, A. R., Sader, J. E. & Roukes, M. L. Nanomechanical mass measurements through feature-based time series clustering, *Rev. Sci. Instrum.* **95,** 025001 (2024).

30. Burg, T. P. & Manalis, S. R. Suspended microchannel resonators for biomolecular detection. *Appl. Phys. Lett.* **83,** 2698–2700 (2003).

31. Burg, T. P., Godin, M., Knudsen, S. M., Shen, W., Carlson, G., Foster, J. S., Babcock, K. & Manalis, S. R. Weighing of biomolecules, single cells and single nanoparticles in fluid. *Nature* **446,** 1066–1069 (2007).

32. Stockslager, M. A., Olcum, S., Knudsen, S. M., Kimmerling, R. J., Cermak, N., Payer, K. R., Agache, V. & Manalis, S. R. Rapid and high-precision sizing of single particles using parallel suspended microchannel resonator arrays and deconvolution. *Rev. Sci. Instrum.* **90,** 085004 (2019).

33. Collis, J. F., Olcum, S., Chakraborty, D., Manalis, S. R. & Sader, J. E. Measurement of Navier slip on individual nanoparticles in liquid. *Nano Lett.* **21,** 4959–4965 (2021).

34. Sanz-Jiménez, A., Ruz, J. J., Gil-Santos, E., Malvar, O., García-López, S., Kosaka, P. M., Cano, Á., Calleja, M. & Tamayo, J. Square membrane resonators supporting degenerate modes of vibration for high-throughput mass spectrometry of single bacterial cells. *ACS Sens.* **8,** 2060–2067 (2023).

35. Ghadimi, A. H., Fedorov, S. A., Engelsen, N. J., Bereyhi, M. J., Schilling, R., Wilson, D. J. & Kippenberg, T. J. Elastic strain engineering for ultralow mechanical dissipation. *Science* **360,** 764–768 (2018).

SUPPORTING INFORMATION

# Data-driven fingerprint nanoelectromechanical mass spectrometry

John E. Sader[1], Alfredo Gomez[2], Adam P. Neumann[2], Alexander R. Nunn[2] and Michael L. Roukes[2]

[1]*Graduate Aerospace Laboratories and Department of Applied Physics, California Institute of Technology, Pasadena, California 91125, USA*
[2]*Department of Physics, California Institute of Technology, Pasadena, California 91125, USA*

**CONTENTS**

A. Mass uncertainty due to database discreteness

B. Variance in the position discrepancy of the analyte

C. Database filtering using noise threshold

D. Numerical validation using a cantilevered beam

E. Experimental validation using polystyrene particles measured with a suspended microchannel resonator

# Section A – Mass uncertainty due to database discreteness

The discrete nature of the fingerprint database introduces uncertainty into mass measurements because the most parallel fingerprint vector extracted from the database may not be truly parallel to the measurement. Here, we quantify this error for 1D mass placement and derive formulas for the resulting uncertainty in the measured mass.

A fingerprint database is said to be "single-valued" if each of its fingerprints has only one direction and magnitude, as its number of elements, $N_{\text{database}} \to \infty$; barring fingerprint sets of zero measure. Such a database has the following properties:

i. A single-valued database can have fingerprints that originate from multiple spatial positions on the device, e.g., due to device symmetry. This does not affect mass measurement accuracy or the developed theory.

ii. Finite $N_{\text{database}}$ can introduce ambiguity in selection of the most parallel vector if the measured fingerprint for the analyte does not belong to the database, as we shall discuss.

The scenario in (ii) can generate large mass anomalies and is excluded from the theory, which is derived in the asymptotic limit of large $N_{\text{database}}$.

**1. Theory for mass uncertainty**

Consider a one-dimensional device, whose governing equation for the fractional frequency shift, $\Delta f_n$, of its eigenmode, $n$, is

$$\Delta f_n = -\frac{M_{\text{analyte}}}{M_{\text{device}}} \Phi_n^2(x), \tag{A1}$$

where $\Phi_n(x)$ is the corresponding displacement field of the eigenmode; $n = 1, 2, \ldots, N$; the analyte mass is $M_{\text{analyte}}$; the device mass is $M_{\text{device}}$; and $x$ is the 1D spatial position of the analyte mass on the device. The fingerprint is defined as the sequence of fractional frequency shifts, i.e., $\{\Delta f_1, \Delta f_2, \ldots, \Delta f_N\}$.

The algorithm chooses a fingerprint in the discrete database that is *most parallel* to the measured fingerprint of the analyte. This is equivalent to choosing the spatial position on the device from which that specific *most parallel vector* arises. Inevitably, there can be a (small) difference between that spatial position and the actual analyte position, $x$, due to database discretization; this difference is denoted $\Delta x$. The components of the most parallel vector are then

$$\Delta f_n|_{\text{database}} = -\frac{M_{\text{database}}}{M_{\text{device}}} \Phi_n^2(x + \Delta x), \tag{A2}$$

where $M_{\text{database}}$ is the standardized mass used to generate the database.

We assume the database has sufficient discretization so that $\Delta x$ can be considered small, i.e., $N_{\text{database}} \gg 1$, and Eq. (A2) is expanded accordingly,

$$\Delta f_n|_{\text{database}} = -\frac{M_{\text{database}}}{M_{\text{device}}} \left( \Phi_n^2(x) + \Delta x\, \Phi_n^{2\,\prime}(x) + \cdots \right), \tag{A3}$$

where $'$ denotes the spatial derivative. Retaining the leading-order terms only and using Eq. (A1), Eq. (A3) becomes

$$\Delta f_n|_{\text{database}} = \frac{M_{\text{database}}}{M_{\text{analyte}}} \Delta f_n|_{\text{true}} - \frac{M_{\text{database}}}{M_{\text{device}}} \Delta x \, \Phi_n^{2\,\prime}(x). \tag{A4}$$

Here, $\Delta f_n|_{\text{true}}$ are components of the analyte's fingerprint at its true position, $x$, i.e., the measured fingerprint from Eq. (A1).

The measured mass of the analyte, using the most parallel vector extracted from the database in Eq. (A4), is

$$M_{\text{measured}} = \frac{\|\{\Delta f_n|_{\text{true}}\}\|}{\|\{\Delta f_n|_{\text{database}}\}\|} M_{\text{database}}, \tag{A5}$$

where $\{\ldots\}$ defines a sequence, i.e., a fingerprint. Substituting Eq. (A4) into Eq. (A5) gives

$$M_{\text{measured}} = \frac{\|\{\Delta f_n|_{\text{true}}\}\|}{\left\|\left\{\frac{M_{\text{database}}}{M_{\text{analyte}}} \Delta f_n|_{\text{true}} - \frac{M_{\text{database}}}{M_{\text{device}}} \Delta x \, \Phi_n^{2\,\prime}(x)\right\}\right\|} M_{\text{database}}, \tag{A6}$$

showing that $M_{\text{measured}}$ can differ from the true analyte mass, $M_{\text{analyte}}$.

Expanding Eq. (A6) to leading order in $\Delta x$, and using Eq. (A1), gives the relative error in the measured mass,

$$\Delta M \equiv \frac{M_{\text{measured}} - M_{\text{analyte}}}{M_{\text{analyte}}} = -\frac{\sum_{n=1}^{N} \Phi_n^2(x) \, \Phi_n^{2\,\prime}(x)}{\sum_{n=1}^{N} \Phi_n^4(x)} \Delta x. \tag{A7}$$

It then follows from Eq. (A7) that the relative variance in the measured mass of an analyte at a *fixed* position, $x$—derived by selecting the most parallel database vector to its *fixed* fingerprint—is

$$\text{Var}\,\Delta M = \frac{1}{4}\left(\frac{1}{F}\frac{dF}{dx}\right)^2 \text{Var}\,\Delta x, \tag{A8}$$

where

$$F(x) = \sum_{n=1}^{N} \Phi_n^4(x), \tag{A9}$$

which is the squared magnitude of a dimensionless fingerprint. Thus, $\text{Var}\,\Delta M$ in Eq. (A8) is proportional to the relative rate-of-change in the squared magnitude of the fingerprint. $\text{Var}\,\Delta x$ is the value over all realizations of the database.

An expression for $\text{Var}\,\Delta x$ is required to complete the solution. The fingerprint database is constructed in the learning phase by randomly placing the mass standard at discrete spatial positions in the interval, $x \in [x_{\min}, x_{\max}]$. As shown in Section B, the analyte's position discrepancy, $\Delta x$, relative to the position corresponding to the most parallel vector, satisfies

$$\text{Var}\, \Delta x = \frac{\delta x^2}{2}, \tag{A10}$$

where

$$\delta x \equiv \frac{x_{\max} - x_{\min}}{N_{\text{database}}}. \tag{A11}$$

Substituting Eqs. (A10) and (A11) into Eq. (A8), gives the variance in the measured mass placed at a single (known) position, $x$, due to selection of the most parallel vector over all realizations of the database,

$$\text{Var}\, \Delta M|_x = \frac{1}{8}\left(\frac{x_{\max} - x_{\min}}{N_{\text{database}}}\right)^2 \left(\frac{1}{F}\frac{dF}{dx}\right)^2. \tag{A12}$$

Next, we consider an ensemble of measurements of the same analyte placed randomly over a finite spatial region of the device, $x \in [x_{\min}, x_{\max}]$. Averaging Eq. (A12) over all analyte positions—because these positions are distributed randomly with uniform probability along the device during mass measurement (see above)—gives the average relative mass uncertainty,

$$\text{Var}\, \Delta M = \frac{x_{\max} - x_{\min}}{8 N_{\text{database}}^2} \int_{x_{\min}}^{x_{\max}} \left(\frac{1}{F}\frac{dF}{dx}\right)^2 dx. \tag{A13}$$

## 2. Mass placement near a stationary support of the device

We consider two practical supports: (A) clamped, i.e., zero displacement and slope; and (B) simply supported, zero displacement and moment.

**Case A:** Near a clamp (at $x = 0$), $\Phi_n(x) \sim a_n x^2$ where $a_n$ is a constant, and Eq. (A13) gives

$$\text{Var}\, \Delta M \sim 2 \frac{x_{\max} - x_{\min}}{x_{\min} N_{\text{database}}^2} \quad \text{as} \quad x_{\min} \to 0. \tag{A14}$$

This establishes that uncertainty in the measured mass diverges if any mass position approaches a clamped support.

**Case B:** For a device that is simply supported at $x = 0$, we have $\Phi_n(x) \sim a_n x$, and Eq. (A13) becomes

$$\text{Var}\, \Delta M \sim \frac{x_{\max} - x_{\min}}{2 x_{\min} N_{\text{database}}^2} \quad \text{as} \quad x_{\min} \to 0, \tag{A15}$$

which also diverges in the same fashion as for a clamped support. This is because the eigenmodes have similar functional forms near the supports, which leads to deterioration of the level of uniqueness in the fingerprint database. Namely, the fingerprints generated near a support are nearly parallel.

This shows that mass placement near the supports of a device should be avoided, not only because of inevitably poor signal-to-noise in the measured fingerprints, but because mass uncertainty diverges. An algorithm for the rejection of these fingerprints is detailed in Section C.

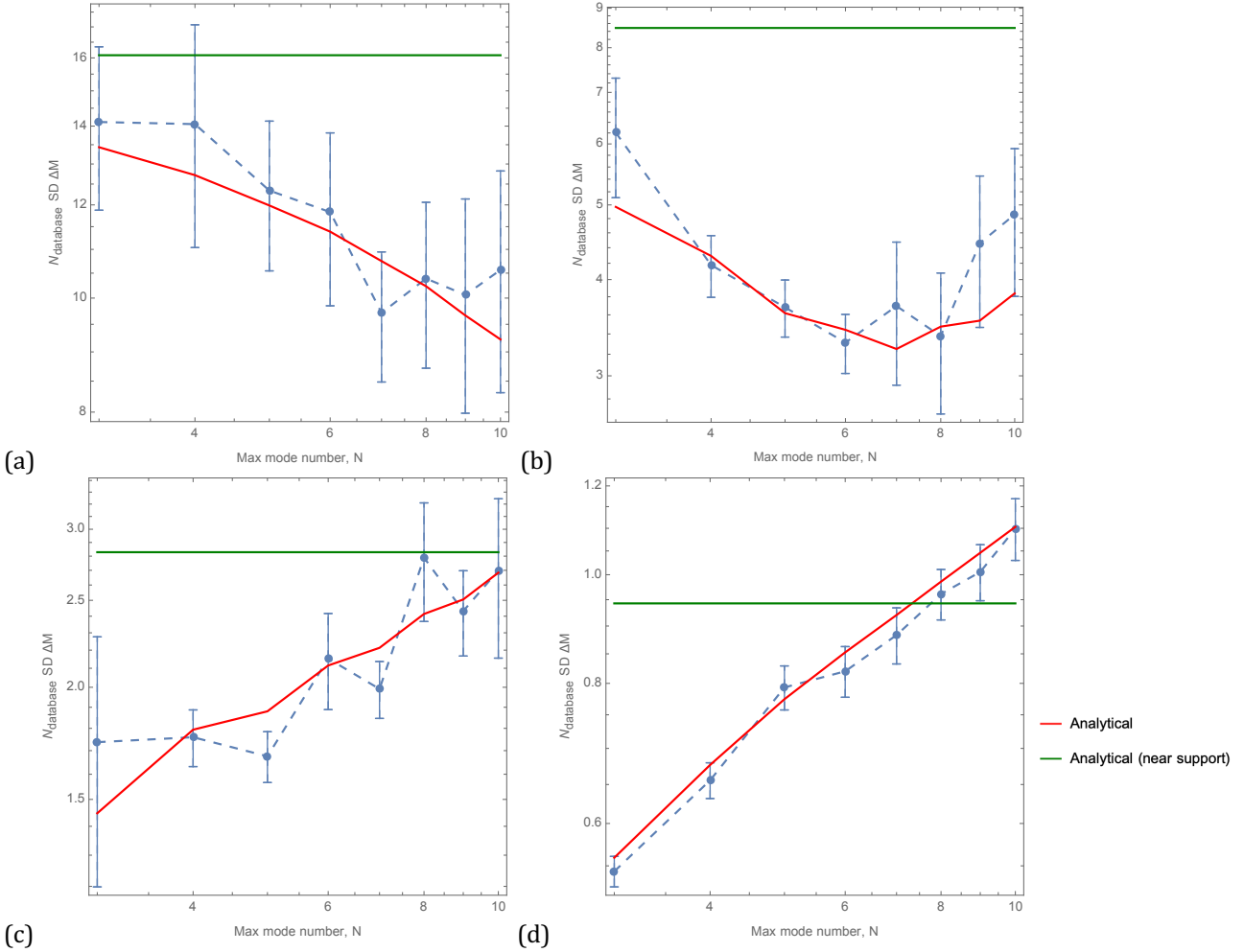

**Figure A1:** Normalized standard deviation of the measured mass, $N_{\text{database}}$ SD $\Delta M$, due to the discrete nature of the fingerprint database. Analyte position range is $x_{\min} < x < 1$: (a) $x_{\min} = 0.03$; (b) $x_{\min} = 0.1$; (c) $x_{\min} = 0.5$; (d) $x_{\min} = 0.9$. Results using (i) near-support formula, Eq. (A14) (green line), (ii) exact formula, Eq. (A13) (red line), and (iii) Monte Carlo simulations with error bars specifying 95% C.I. Number of fingerprints in the database is indicated. $N_{\text{database}} = 100$; similar results for $N_{\text{database}} = 300, 1000$ (data not shown). Mass measurements are performed 100 times, which are repeated 100 times to estimate 95% confidence intervals. Outliers discussed in Section 4 are rejected by only accepting simulations where the mean and median of the repeat 100 samples (to compute the 95% C.I.) differ by no more 50%.

## 3. Comparison to Monte Carlo simulations

Figure A1 compares Eqs. (A13) and (A14) to Monte Carlo simulations for ensemble measurements of a single analyte over a finite region of a cantilevered beam, employing at least 3 eigenmodes—satisfying the database single-valued requirement (see definition above). The distance of the minimum analyte position to the clamped end, $x_{\min}$, is varied to illustrate the effect of mass measurement near a support (clamp). The results in Figure A1 validate Eq. (A13) and show that increasing the number of eigenmodes from $N = 3$ to 10 has a weak effect on the mass measurement uncertainty. Interestingly, the near support formula, Eq. (A14), gives reasonable performance for all $x_{\min}$, estimating the true value to within a factor of approximately two.

## 4. Outliers in mass measurement due to database discreteness

Finite (and large) $N_{database}$ can produce significant error in the measured mass even when the database is single valued. This is because the algorithm can choose a most parallel vector from the (discrete) database that is far from the true solution. Namely, the true solution may be missing from the database due to its discreteness, and the most parallel vector may be on a different branch of the locus-of-points of the database. Consequently, the theory may not give the correct mass and either higher $N_{database}$ or a greater number of eigenmodes, $N$, may be needed.

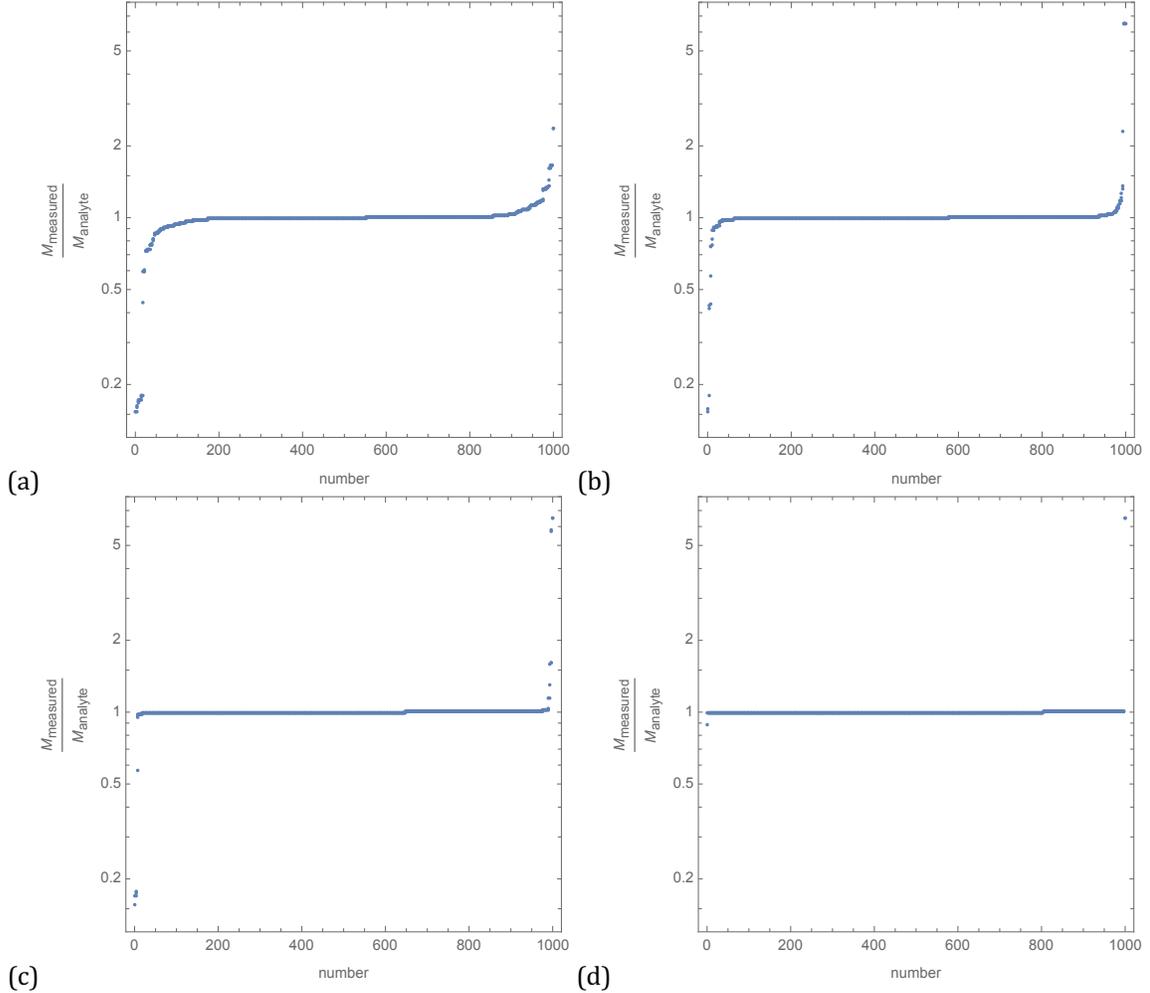

**Figure A2:** Sorted density plots of 1,000 measured scaled values of the analyte mass, $M_{measured}/M_{analyte}$, for random placement on a cantilevered beam along $x \in [0.1, 0.9]$, with (a) $N_{database} = 300$, (b) $N_{database} = 1,000$, (c) $N_{database} = 3,000$, (d) $N_{database} = 10,000$. Using eigenmodes $n = 3, 4, 5$ only, which gives a single-valued fingerprint database (see definition above). Large outliers arise from selection of a most parallel vector that lies far from the true solution, due to finite $N_{database}$. The number of outliers is observed to vary as $1/N_{database}$, i.e., identical to SD $\Delta M$ in Eq. (A13). Their overall contribution to SD $\Delta M$ is undiminished by increasing $N_{database}$, even though the overall value of SD $\Delta M$ decreases.

An example of this effect is shown in Fig. A2 for a cantilevered beam with analyte positions in the region $x \in [0.1, 0.9]$ that uses eigenmodes $n = 3$ to 5, which gives a single-valued fingerprint database (because it employs 3 eigenmodes; see above). The sorted density plot in Fig. A2 shows outliers that deviate strongly from the true mass. Removing these outliers (choosing a threshold of 10% from the mean) gives sorted density plots that resemble Gaussians distributions and values for SD $\Delta M$ that agree with Eq. (A13); see Table A1. Increasing the number of eigenmodes used to 4 (i.e., $n = 3$ to 6) eliminates the above effect, greatly reduces the occurrence of outliers, and gives good agreement with Eq. (A13) (data not shown).

**Table A1:** Normalized standard deviation of the uncertainty in the measured mass, $N_{\text{database}}$ SD $\Delta M$, due to the discrete nature of the fingerprint database, and the effect of outliers (with their removal). Results shown for a cantilevered beam with mass placement on $x \in [0.1, 0.9]$, using eigenmodes $n = 3, 4, 5$.

| $N_{\text{database}}$ | $N_{\text{database}}$ SD $\Delta M$ | | |
|---|---|---|---|
| | Eq. (A13) | Monte Carlo | Monte Carlo (outliers removed) |
| 100 | | 18.32 | 3.612 |
| 300 | 5.353 | 27.39 | 5.643 |
| 1,000 | | 56.07 | 5.330 |
| 3,000 | | 57.91 | 4.928 |

Interestingly, we do not observe this effect if eigenmodes $n = 1$ to $3$ are used. This is because the locus-of-points of the fingerprint database is simpler in this case, which reduces the possibility of selecting a most parallel vector (from different branches) that gives an incorrect mass; see Fig. A3.

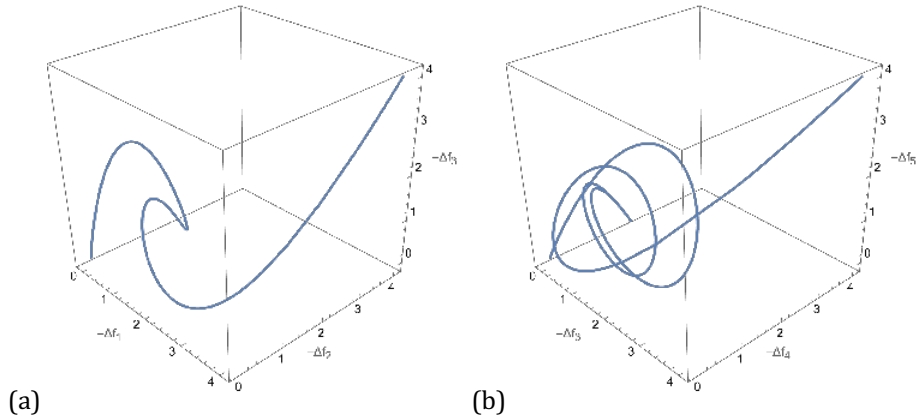

(a)        (b)

**Figure A3:** Loci-of-points of fingerprint databases for a cantilevered beam, when using three eigenmodes: (a) $n = 1, 2, 3$; (b) $n = 3, 4, 5$. The relative frequency shifts are normalized by $M_{\text{analyte}}/2M_{\text{device}}$.

## 5. Effect of device eigenmode symmetries

If the device exhibits symmetries, e.g., the doubly-clamped beam discussed above, it can change the distance between the spatial positions from which the database fingerprints arise and modify the analysis. Randomness in position selection generates unique fingerprints in each symmetric region—there is no duplication. This reduces the effective distance between the spatial positions of the database fingerprints—by the number of symmetric regions, $N_{\text{multi}}$—when mapped onto the original symmetric region used in Eq. (A13). That is, the variance, Var $\Delta x$, in Eq. (A12) is reduced by a factor of $N_{\text{multi}}^2$ and hence Eq. (A13) becomes

$$\text{Var}\,\Delta M = \frac{x_{\max} - x_{\min}}{8 N_{\text{database}}^2 N_{\text{multi}}^2} \int_{x_{\min}}^{x_{\max}} \left(\frac{1}{F}\frac{dF}{dx}\right)^2 dx. \tag{A16}$$

Monte Carlo simulations (not shown) confirm the validity of Eq. (A16).

## Section B – Variance in the position discrepancy of the analyte

Let $Y$ and $\{X_i\}_{i=1}^n$ be independent random variables with uniform distribution in the unit interval, i.e., $U(0,1)$. The displacement of $Y$ from the random grid $\{X_i\}_{i=1}^n$ is defined as the analyte's position discrepancy (see Section A), and is given by the random variable,

$$\Delta x = \mathrm{argmin}_{Y-X_i}|Y - X_i|. \tag{B1}$$

It then follows that $\Delta x$ has expectation, $E[\Delta x] = 0$, and variance:

$$\mathrm{Var}\,\Delta x = \frac{n+7}{2(n+1)(n+2)(n+3)} \sim \frac{1}{2n^2} \text{ as } n \to \infty, \tag{B2}$$

where $n$ is the number of fingerprints in the database, i.e., $n = N_{\mathrm{database}}$; Eq. (B2) is used in Eq. (A10).

**Proof**

The expectation, $E[\Delta x^k]$, is given by

$$E[\Delta x^k] = \sum_{i=1}^n E[(Y - X_i)^k \cdot \mathbb{1}\{|Y - X_i| < \min_{j \neq i}|Y - X_j|\}], \tag{B3}$$

which can be rewritten as

$$E[\Delta x^k] = n E[(Y - X_1)^k \cdot P(|Y - X_1| < |Y - X_2|)^{n-1}], \tag{B4}$$

where listed probability in Eq. (B4) is

$$P(|Y - X_1| < |Y - X_2| \mid X_1 = x_1, Y = y) = \max[0, 1 - y - |x_1 - y|] + \max[0, y - |x_1 - y|]. \tag{B5}$$

Equation (B5) is easily evaluated, to give

$$P(|Y - X_1| < |Y - X_2| \mid X_1 = x_1, Y = y) = \begin{cases} x_1 & : \dfrac{x_1+1}{2} < y \\ 1 + 2x_1 - 2y & : x_1 < y < \dfrac{x_1+1}{2} \\ 1 + 2y - 2x_1 & : \dfrac{x_1}{2} < y < x_1 \\ 1 - x_1 & : 0 < y < \dfrac{x_1}{2} \end{cases}, \tag{B6}$$

Substituting Eq. (B6) into Eq. (B4), we obtain the integral expression,

$$E[\Delta x^k] = n \int_0^1 dx_1 \left\{ \int_{(x_1+1)/2}^1 (y - x_1)^k x_1^{n-1} dy + \int_{x_1}^{(x_1+1)/2} (y - x_1)^k (1 + 2x_1 - 2y)^{n-1} dy \right.$$
$$\left. + \int_{x_1/2}^{x_1} (y - x_1)^k (1 + 2y - 2x)^{n-1} dy + \int_0^{x_1/2} (y - x_1)^k (1 - x_1)^{n-1} dy \right\}. \tag{B5}$$

which gives

$$E[\Delta x] = 0, \tag{B6a}$$

$$E[\Delta x^2] = \frac{n+7}{2(n+1)(n+2)(n+3)}. \tag{B6b}$$

It then follows that the required variance is

$$\text{Var } \Delta x = E[\Delta x^2] - E[\Delta x]^2 = \frac{n+7}{2(n+1)(n+2)(n+3)}, \tag{B7}$$

which agrees with Eq. (B2).

## Section C – Database filtering using noise threshold

In this section, we develop an algorithm to reject fingerprints that arise from analyte adsorption close to the clamped ends of a NEMS device, to minimize uncertainty due to database discreteness. This is achieved by imposing a noise-based threshold.

A dataset of noise measurements can always be generated by measuring the NEMS sensor response without analyte adsorption. From this noise dataset, we construct a fingerprint noise database, denoted $F_{\text{noise}}$, that contains the fingerprints, $\Omega_{\text{noise}}$. We consider NEMS noise measurements to follow a multivariate normal distribution. The contours of constant probability density are then given by vectors, $x$, satisfying:

$$(x - \mu)^{\text{T}} \Sigma^{-1} (x - \mu) = c^2, \tag{C1}$$

where $\mu$ is the mean vector, $\Sigma$ is the covariance matrix and $c^2$ is the confidence region. To define confidence regions, $c^2$ can be set to the quantiles of the $\chi^2$ distribution. For example, for 95% confidence values, one can use the inverse $\chi^2$ for a $p$-value of 0.05 with the degrees-of-freedom equal to the number of vector dimensions (i.e., number of eigenmodes).

**Example:** For the SMR measurements reported in Section 4B and Supplementary Information Section E, the vector dimension is 5, which for example gives $c^2$ =11.07 when assuming a 95% C.I. Measured fingerprints can be accepted (or rejected) based on whether they lie outside (or inside) this chosen confidence interval. For example, points can be classified as noise (and rejected) if the left-hand-side of Eq. (C1) is less than or equal to $c^2$, or conversely, classified as accepted if the left-hand-side of Eq. (C1) is greater than $c^2$. The effect of such filtering, using different choices of quantiles of the $\chi^2$ distribution, is shown in Figures C1 and C2.

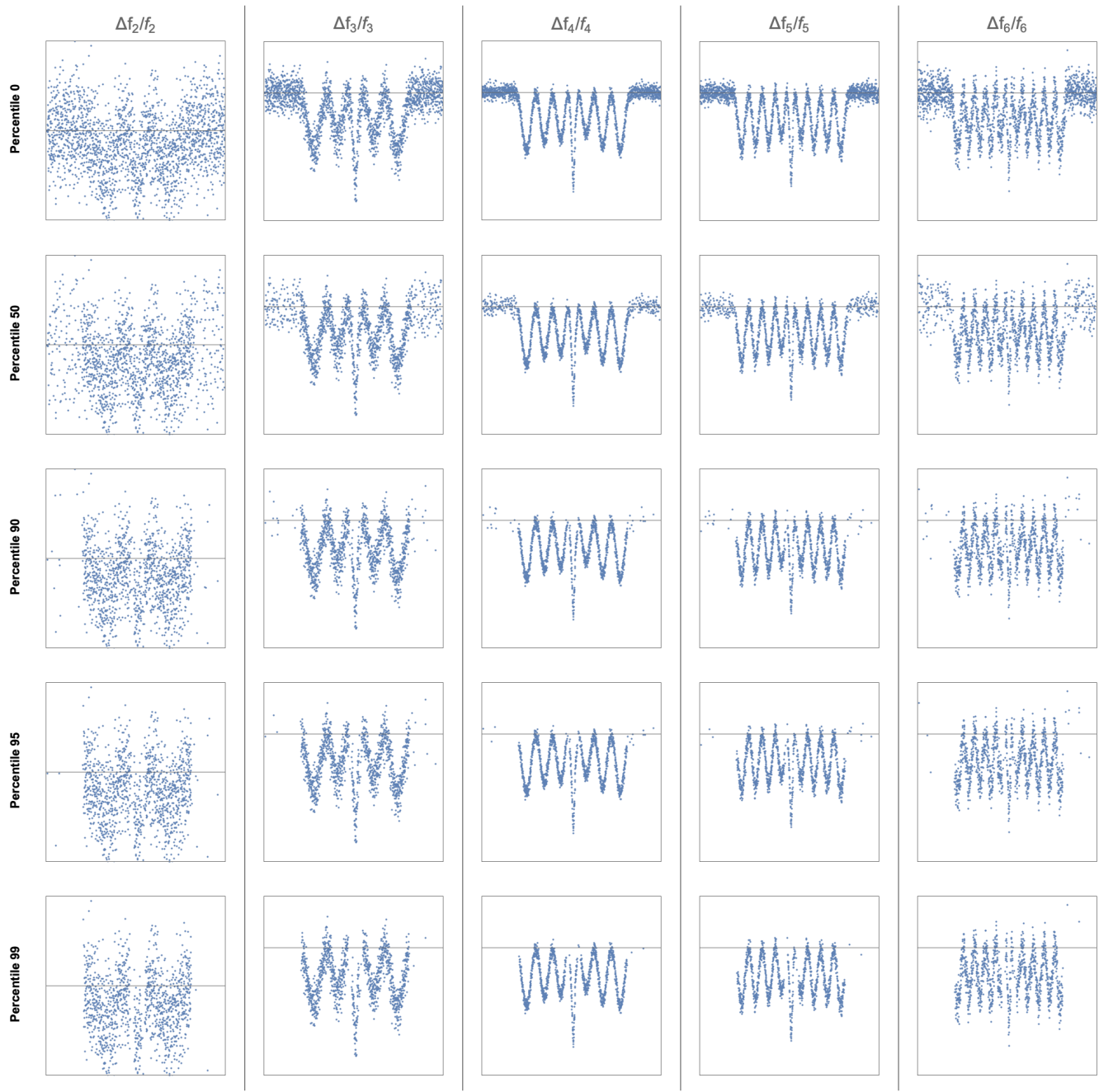

**Figure C1:** Filtered time series of the relative frequency shifts of eigenmodes 2 to 6 for varying noise thresholds, as an analyte particle moves through the SMR. The noise threshold, given by the percentile confidence, begins at $0^{th}$ percentile (raw time-series) and increases from the top to bottom row. Increasing the percentile confidence to a level of 95% is observed to exclude fingerprints near the clamp.

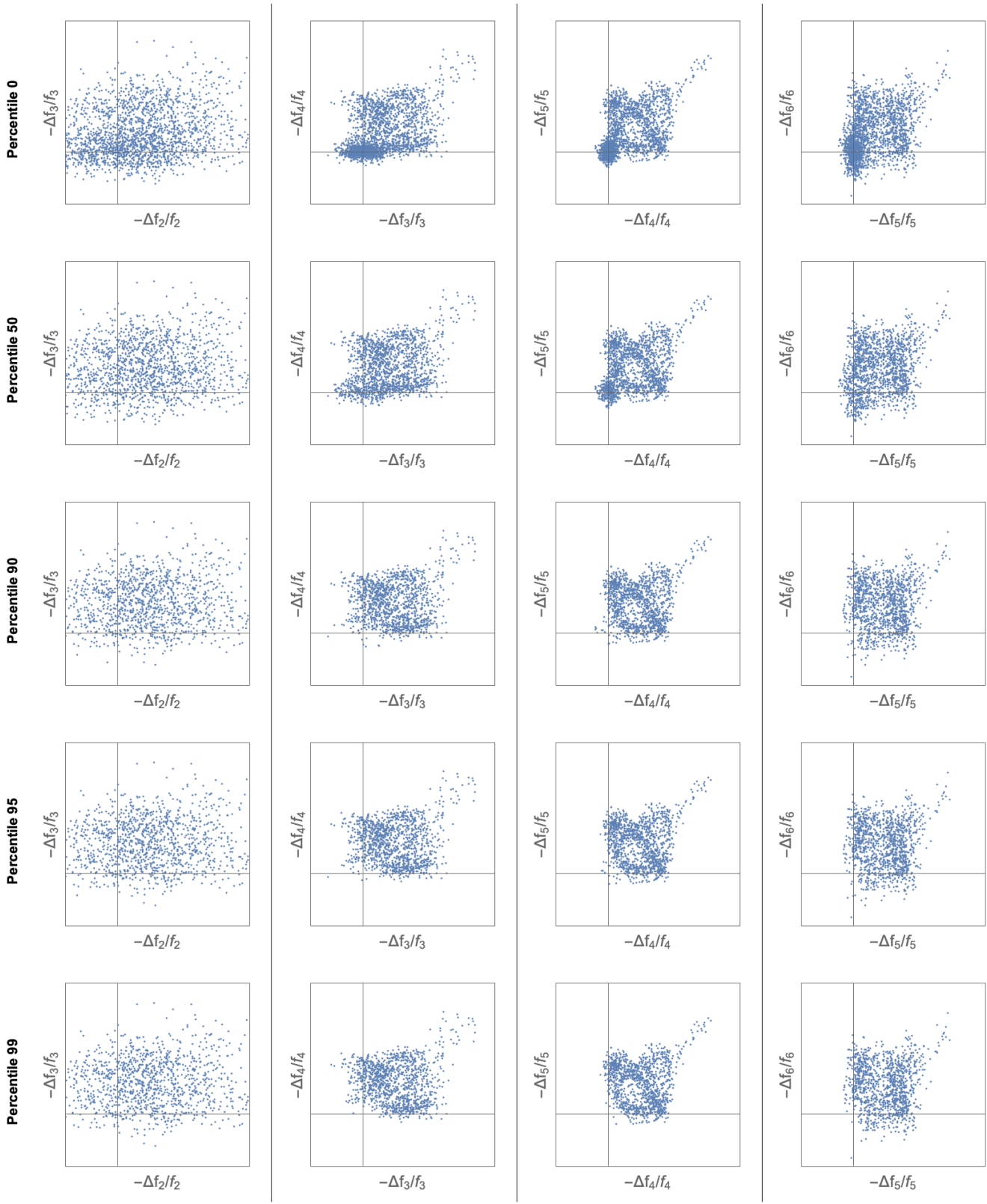

**Figure C2:** Configuration space of eigenmode pairs, where noise filtering is applied (as per Fig. C1). This filtering is observed to exclude fingerprints near the origin, which correspond to analytes near the clamp.

## Section D – Numerical validation using a cantilevered beam

In this section, we present a numerical validation of the fingerprint method for a (one-dimensional) cantilevered beam. This complements the numerical validation reported in Section 3 for a multi-dimensional advanced NEMS device. Following a similar procedure as outlined in Section 3, particles are placed along a one-dimensional cantilever. Using the first two flexural eigenmodes only gives the correct mass with a standard deviation of ~30% over all measurements, while use of the first three eigenmodes exhibits a standard deviation of only ~0.6%.

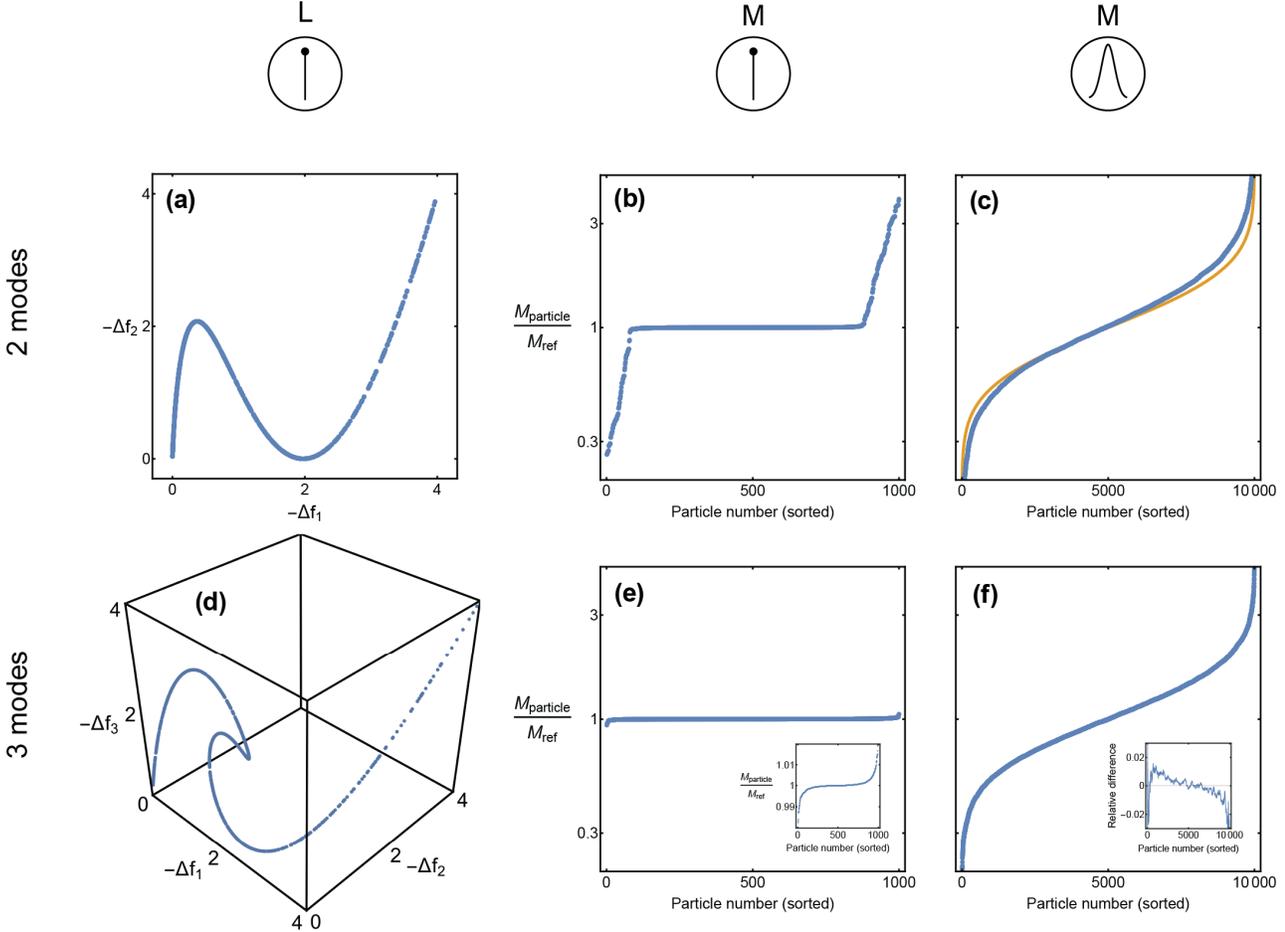

**Figure D1:** Fingerprint approach for a (one-dimensional) cantilevered beam with mass adsorption occurring at random position on the interval, $0.1 < x < 1$. The cantilever's clamped and free ends are at $x = 0$ and 1, respectively. Results shown for the first two and three flexural eigenmodes. **(a, d)** Learning phase showing fingerprint databases with $N_{\text{database}} = 1{,}000$ fingerprints, which are used in all measurement phases. **(b, e)** Measurement phases for 1,000 identical analyte particles, shown as an ordered density plot. **(c, f)** Measurement phases for 10,000 different analyte particles, whose masses obey a log-normal distribution. Orange curve is the true mass distribution. Blue dots (which appear as a single curve) are the measured masses using the fingerprint approach. Inset in (f) shows the relative difference between the true and measured mass distributions; it is at 1% level. The number of analyte particles is unlimited and independent of the fingerprint database size.

Regardless of the algorithm used to interpret the frequency shifts, unique mass measurement on a one-dimensional device is guaranteed only provided at least three eigenmodes are used (for a one-dimensional device). The poorer accuracy exhibited in the two-eigenmode data is due to the (non-unique) multivalued nature of its fingerprint database; see Fig. D1(a). This can and does produce strong outliers in the measured mass when the wrong branch of solution is selected by Eq. (3). Even so, the measurement phase is observed to determine the correct mass with a probability of ~80%.

Increasing the number of eigenmodes to at least three generates a single-valued database in the learning phase, in contrast to the multiple solution branches of the 2-eigenmode database. This property of the 3-eigenmode learning phase is true for all NEMS devices where the effect of particle position on the fingerprint is restricted to a one-dimensional space. This is because the loci-of-points of $\mathbf{\Omega}_{\text{database}} \in \mathbf{F}$—which are parametrized by the one-dimensional particle position along the device—are curves in the fingerprint database's $N$-dimensional configuration space, e.g., see Fig. D1(d). When three or more eigenmodes are used, any ambiguity in the selection of $\mathbf{\Omega}_{\text{parallel}}$ will have a probability of zero.[†]

Figures D1(c, f) report mass measurements on the cantilevered beam for a set of 10,000 analyte particles of different masses satisfying a log-normal distribution in the measurement phase. The results are analogous to those present in Section 3 of the text. Moreover, changing the number of eigenmodes adjusts this error, which is detailed in Supplementary Information Section A.

**DIMENSIONALITY:** At least four eigenmodes are required to guarantee single-valuedness for two-dimensional adsorption on the physical surfaces of NEMS devices, because the effect of particle position on the fingerprint is parametrized by two independent variables.

---

[†] This property is strictly true in the limit, $N_{\text{database}} \to \infty$.

# Section E – Experimental validation using polystyrene particles measured with a suspended microchannel resonator

We study multimode mass measurements of polystyrene particles in water, using a suspended microchannel resonator (SMR). The SMR is a microcantilever with an embedded microfluidic channel within which fluid transports the analyte. Simultaneous measurements of the resonant frequencies of flexural eigenmodes 2 to 6 are taken as individual polystyrene particles passed through the SMR. The fingerprint approach (which involves no model) is compared to independent analysis that fits the Euler-Bernoulli beam theory eigenmodes to the measured frequency shifts. These SMR measurements and their details were reported in Collis *et al.*[1]

The SMR used in this study and its measurement setup is summarized in Fig. E1 and Table E1, with details provided in Collis *et al.*[1] NIST-tracible polystyrene particles (ThermoFisher 4016A) were measured and have a characterized radius of 793.5±9(SD) nm with a mass density of 1,050 kg/m$^3$, i.e., the particle mass exhibits a standard deviation of 3%. Resonant frequencies of SMR flexural eigenmodes 2 to 6 were simultaneously measured in real-time using a series of PLL circuits. A sample of the resulting frequency shift time series of flexural eigenmodes 2 to 6 is provided in Fig. E2 along with fits to Euler-Bernoulli beam theory for each eigenmode. The signal-to-noise ratio for eigenmode 2 is observed to be the poorest.

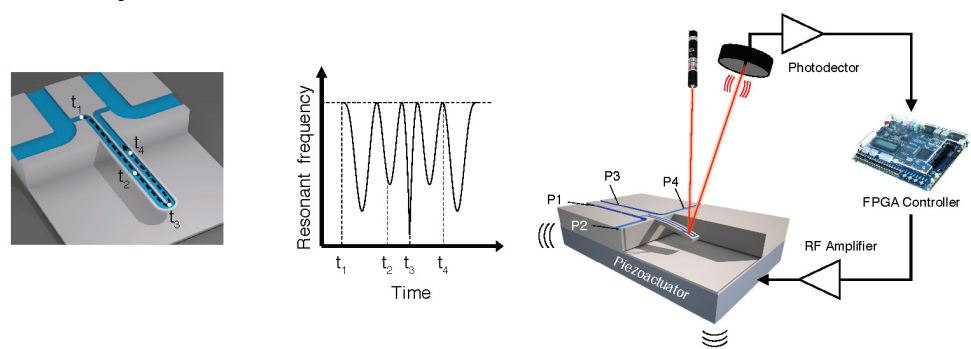

**Figure E1:** Schematic of SMR used in this study. Left panel: SMR as an individual particle passes through its embedded microfluidic channel. Middle panel: Resulting measured resonant frequency of SMR as a function of time, with individual times referencing left panel. Right panel: Measurement setup showing actuation and detection scheme. Taken from Collis *et al.*[1] where all details can be found.

**Table E1:** Properties of SMR used in this study; taken from Collis *et al.*[1]

| Property | Dimension (µm) | Eigenmode number | Unloaded frequency (MHz) | Quality factor | Standard deviation of frequency noise (Hz) |
|---|---|---|---|---|---|
| SMR length | 400 | | | | |
| SMR width | 19 | | | | |
| SMR thickness | 4 | 2 | 0.2180 | 9,440 | 0.43 |
| Channel height | 3 | 3 | 0.6094 | 4,773 | 0.37 |
| Channel width | 5 | 4 | 1.191 | 2,823 | 0.35 |
| Lid thickness | 0.5 | 5 | 1.963 | 2,093 | 0.61 |
| Wall width | 2 | 6 | 2.921 | 1,620 | 1.6 |

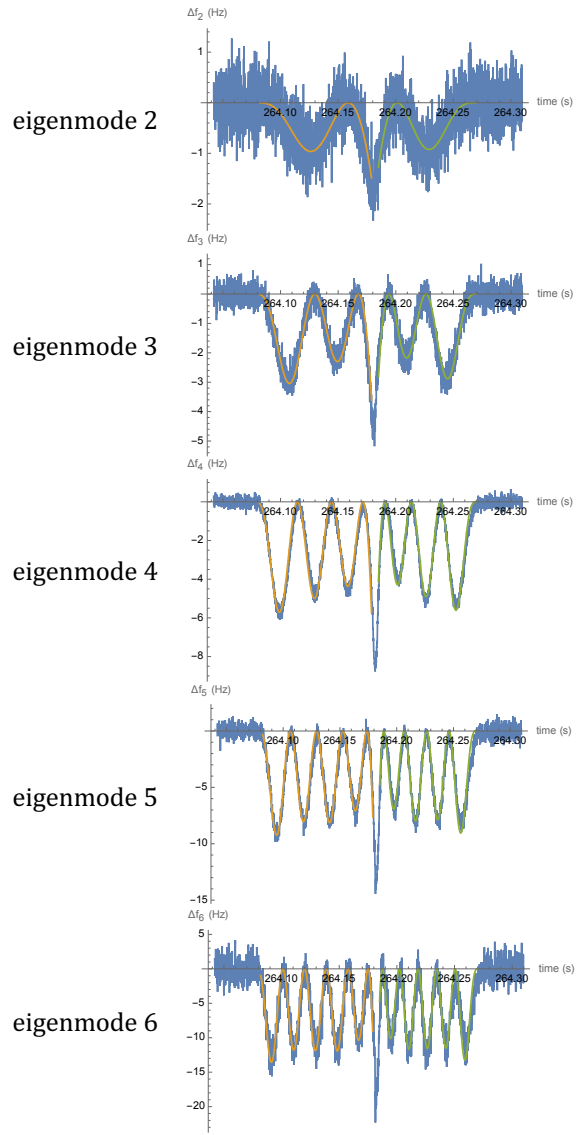

**Figure E2:** Simultaneously measured frequency shifts of flexural eigenmodes 2 to 6 (top to bottom) due to the passage of a single polystyrene particle through the SMR. The particle increases the inertial mass of the SMR which reduces the resonant frequency of each eigenmode. Also shown are best fits to Euler-Bernoulli beam theory eigenmode shapes; these a separately fit to the left- (orange) and right-hand (green) branches of each frequency shift curve, corresponding to passage of the particle through the left- and right-hand microfluidic channels of the SMR.

Fingerprints are constructed using these simultaneous multimode frequency-shift measurements, as described in the text. Figure E3 shows a comparison of the fingerprint database in its corresponding configuration space and Euler-Bernoulli beam theory, for the particle studied in Fig. E2 when only 2 eigenmodes are used. Measured fingerprints show significant noise and display a weak correspondence with theory. However, reducing the noise levels in the measured time series data, by averaging the frequency shift signals amongst neighboring time measurements, leads to strong agreement with theory. This is expected given the agreement between theory and measurement in Fig. E2. Figure E4 shows analogous configuration space plots when 3 eigenmodes are used, where excellent agreement between measurement and Euler-Bernoulli theory is observed upon averaging.

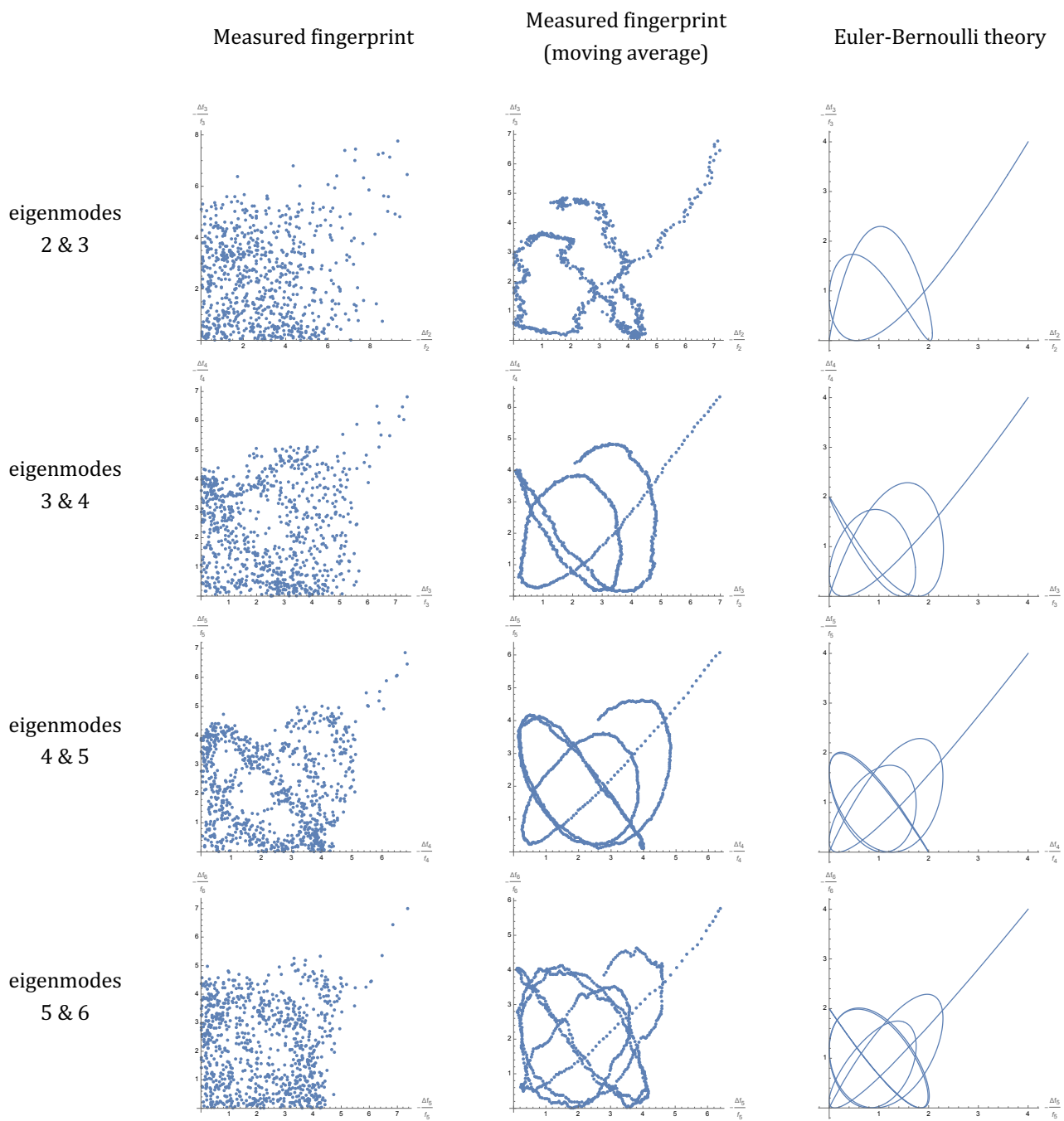

**Figure E3:** Fingerprint database ($\times 10^{-6}$) for a single particle using two flexural eigenmodes. Axes scaled by $10^{-6}$. Comparison of raw measurement (left), measurement smooth using moving average of 20 nearest-neighbors in frequency-shift time-series (middle) and Euler-Bernoulli theory (right). Measurements near the clamp and free end of the SMR are omitted. Measurements taken in left-hand microfluidic channels of the SMR only. Similar results obtained using right-hand channel (data not shown).

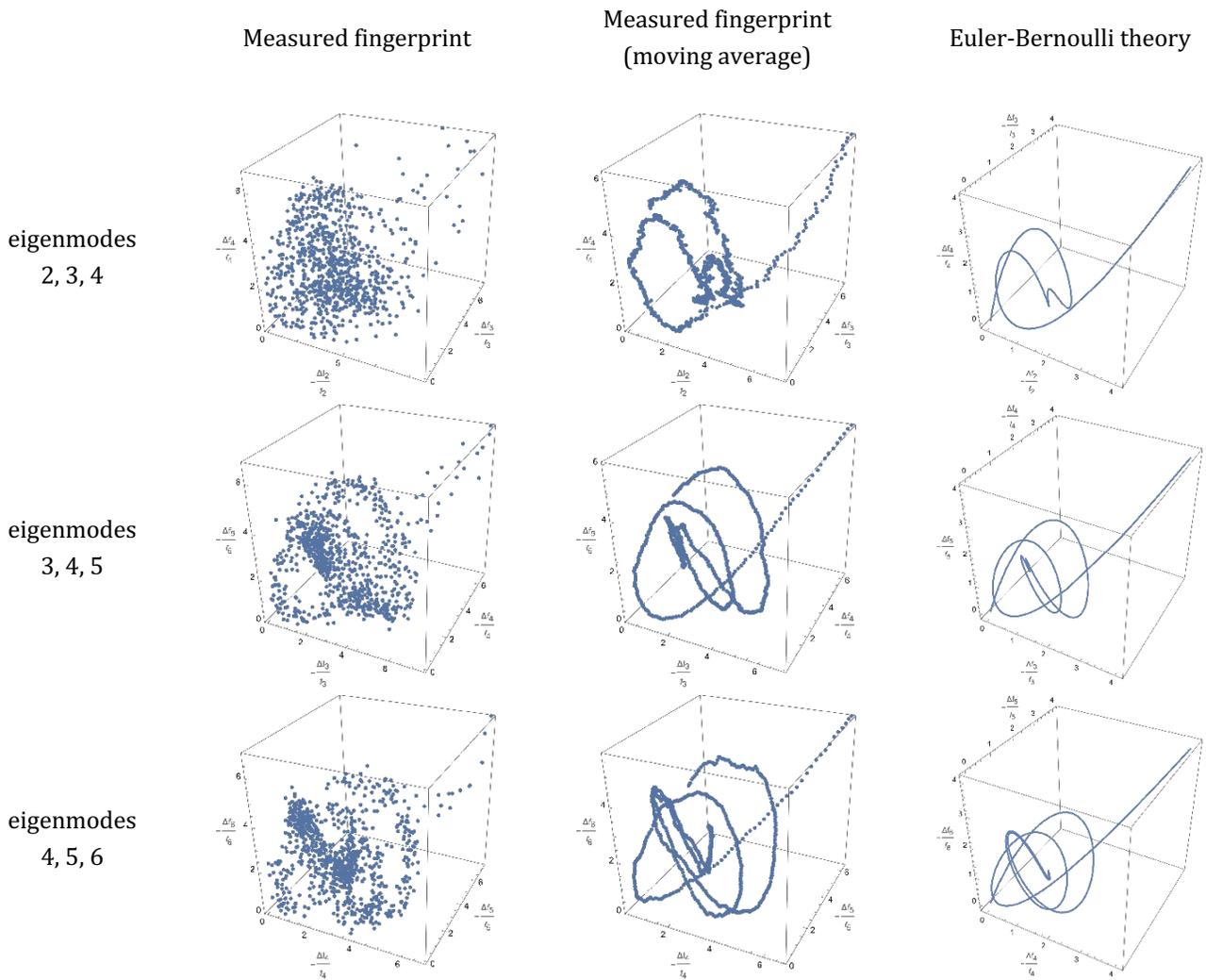

**Figure E4:** As for Fig. E3, but using 3 flexural eigenmodes.

The reported time-series fits to Euler-Bernoulli beam theory in Fig. E2 (with similar fits for the other 340 particles) were used by Collis *et al.*[1] to measure the ratio of the mass of each of the 341 polystyrene particles to the SMR mass (which was unknown), i.e., $M_{\text{particle}}/M_{\text{SMR}}$, for each eigenmode. The NIST standardized mass of the polystyrene particles is 2.197 ± 0.075 (SD) ag, which Collis *et al.*[1] used to measure the SMR mass (not required here). The particles exhibit a minimum "mass discrepancy parameter" of 0.98 for mode 6, establishing that particle motion relative to the SMR is negligible. For example, including the mass discrepancy parameter in the mass measurements of each mode (see Collis *et al.*[1]) leads to a difference in the overall measured mass of only 0.4% relative to the average of the mass obtained from each mode without accounting for the mass discrepancy parameter.

# References

1. Collis, J. F., Olcum, S., Chakraborty, D., Manalis, S. R. & Sader, J. E. Measurement of Navier slip on individual nanoparticles in liquid. *Nano Letters* **21**, 4959–4965 (2021).